\documentclass[fleqn,usenatbib]{mnras}

\usepackage{newtxtext,newtxmath}
\usepackage{color,ulem}

\usepackage[T1]{fontenc}
\usepackage[dvipsnames]{xcolor}
\DeclareRobustCommand{\VAN}[3]{#2}
\let\VANthebibliography\thebibliography
\def\thebibliography{\DeclareRobustCommand{\VAN}[3]{##3}\VANthebibliography}

\newcommand{\Msun}{\ensuremath{\mathrm{M}_\odot}}
\newcommand{\epe}{\ensuremath{\varepsilon_{\rm{e}}}}
\newcommand{\bepe}{\ensuremath{\bar{\varepsilon}_{\rm{e}}}}
\newcommand{\bepen}{\ensuremath{\bar{\varepsilon}_{\rm{e,-1}}}}
\newcommand{\epb}{\ensuremath{\varepsilon_{\rm{B}}}}
\newcommand{\epbn}{\ensuremath{{\varepsilon}_{\rm{B,-1}}}}
\newcommand{\MP}{\ensuremath{m_{\rm{p}}}}
\newcommand{\ME}{\ensuremath{m_{\rm{e}}}}
\newcommand{\SGM}{\ensuremath{\sigma_{\rm{T}}}}

\newcommand{\Eej}{\ensuremath{{E}_{\rm{ej}}}}
\newcommand{\Eejn}{\ensuremath{{E}_{\rm{ej,51}}}}

\newcommand{\tdec}{\ensuremath{{t}_{\rm{dec}}}}
\newcommand{\gamm}{\ensuremath{{\gamma}_{\rm{m}}}}
\newcommand{\num}{\ensuremath{{\nu}_{\rm{m}}}}
\newcommand{\nua}{\ensuremath{{\nu}_{\rm{a}}}}

\newcommand{\Gamin}{\ensuremath{{\Gamma}_{\rm{in}}}}
\newcommand{\betain}{\ensuremath{{\beta}_{\rm{in}}}}
\newcommand{\tnr}{\ensuremath{{t}_{\rm{NR}}}}
\newcommand{\tj}{\ensuremath{{\tau}_{\rm{j}}}}

\newcommand{\Ej}{\ensuremath{{E}_{\rm{j}}}}
\newcommand{\Ejiso}{\ensuremath{{E}_{\rm{j,iso}}}}
\newcommand{\Ejn}{\ensuremath{{E}_{\rm{j,49}}}}
\newcommand{\bdn}{\ensuremath{{\beta}_{\rm{DN}}}}

\newcommand{\tcol}{\ensuremath{{t}_{\rm{col}}}}
\newcommand{\teq}{\ensuremath{{t}_{\rm{eq}}}}

\newcommand{\Mej}{\ensuremath{{M}_{\rm{ej}}}}

\usepackage{graphicx}	
\usepackage{amsmath}	

\title[Radio remnants of short GRBs]{Searching for the radio remnants of short duration gamma-ray bursts}

\author[R. Ricci et al.]{
R. Ricci,$^{1,2}$\thanks{E-mail: ricci@ira.inaf.it (IRA)}
E. Troja,$^{3,4}$
G. Bruni,$^{5}$
T. Matsumoto,$^{6,7,8}$
L. Piro,$^{5}$
B. O'Connor,$^{9,10,4,3}$\newauthor
T. Piran,$^{6}$
N. Navaieelavasani,$^{5,11}$
A. Corsi,$^{12}$ 
B. Giacomazzo,$^{13,14,15}$
M. H. Wieringa$^{16}$
\\
$^{1}$ Istituto Nazionale di Ricerche Metrologiche - Torino, Italy\\
$^{2}$ INAF - Istituto di Radioastronomia - Bologna, Italy \\
$^{3}$ Astrophysics Science 
Division, NASA Goddard Space Flight Center, 8800 Greenbelt Rd, Greenbelt, MD 20771, USA\\ 
$^{4}$ Department of Astronomy, University of Maryland, College Park, MD 20742-4111, USA\\ 
$^{5}$ INAF - Istituto di Astrofisica e Planetologia Spaziali, via Fosso del Cavaliere 100, I-00133 Roma, Italy \\
$^{6}$ Racah Institute of Physics, Edmund J. Safra Campus, Hebrew University of Jerusalem, Jerusalem 91904, Israel \\
$^{7}$Research Center for the Early Universe, Graduate School of Science, University of Tokyo, Tokyo 113-0033, Japan\\
$^{8}$Department of Physics, Graduate School of Science, University of Tokyo, Tokyo 113-0033, Japan\\
$^{9}$ Department of Physics, The George Washington University, 725 21st Street NW, Washington, DC 20052, USA\\ 
$^{10}$Astronomy, Physics, and Statistics Institute of Sciences (APSIS), The George Washington University, Washington, DC 20052, USA \\
$^{11}$ Universit\'{a} La Sapienza, Rome, Italy \\
$^{12}$ Department of Physics and Astronomy, Texas Tech University, Box 1051, Lubbock, TX, 79409-1051, USA \\
$^{13}$ Dipartimento di Fisica G. Occhialini, Universit\`a di Milano - Bicocca, Piazza della Scienza 3, I-20126 Milano, Italy\\
$^{14}$ INFN, Sezione di Milano Bicocca, Piazza della Scienza 3, I-20126 Milano, Italy \\
$^{15}$ INAF, Osservatorio Astronomico di Brera, Via E. Bianchi 46, I-23807 Merate, Italy \\
$^{16}$ CSIRO Astronomy and Space Science, Epping, NSW, Australia \\
}

\date{Accepted XXX. Received YYY; in original form ZZZ}

\pubyear{2015}

\begin{document}
\label{firstpage}
\pagerange{\pageref{firstpage}--\pageref{lastpage}}
\maketitle

\begin{abstract}
Neutron star mergers produce a substantial amount of fast-moving ejecta, expanding outwardly for years after the merger. The interaction of these ejecta with the surrounding medium may produce a weak isotropic radio remnant, detectable in relatively nearby events. 
We use late-time radio observations of short duration gamma-ray bursts (sGRBs) to constrain this model. Two samples of events were studied: four sGRBs  that are possibly in the local (<200 Mpc) universe were selected to constrain the remnant non-thermal emission from the sub-relativistic ejecta,  whereas 17 sGRBs at cosmological distances were used to constrain the presence of a proto-magnetar central engine, possibly re-energezing the merger ejecta. 
We consider the case of GRB~170817A/GW170817, and find that in this case
the early radio emission may be quenched by the jet blast-wave. 
In all cases, for ejecta mass range of $\Mej\lesssim10^{-2}\,(5\times10^{-2})\,\Msun$, we can rule out very energetic merger ejecta $\Eej\gtrsim5\times10^{52}\,(10^{53})\,\rm erg$, thus excluding the presence of a powerful magnetar as a merger remnant.

\end{abstract}




\section{Introduction}

Neutron star (NS) mergers are prime sources of gravitational wave (GW) radiation detectable by the advanced LIGO and Virgo experiments \citep{gw170817,gw190425}, and likely the dominant site for the production of heavy ($A$\,$\gtrsim$160) elements
\citep[e.g.][]{Lattimer74,Eichler89,Freiburghaus99}. 
The recent discovery of GW170817/GRB170817A \citep{gw170817,grb170817,Savchenko2017} confirmed that these systems produce short duration gamma-ray bursts (GRBs): luminous flashes of gamma-ray radiation lasting less than 2 seconds \citep{Kouveliotou1993}. These are followed by a longer lived emission, known as afterglow, visible from radio to X-ray energies \citep{Gehrels2005,Troja2017,Hallinan2017}. 

 In addition to these well-established electromagnetic (EM) counterparts, \citet{np11}  suggested the emergence of a radio flare on timescales of several months/years after the merger. Numerical simulations of NS mergers ubiquitously show that a small fraction of matter ($\lesssim$0.1 $M_{\odot}$) is dynamically ejected during the merger itself, or later in the evolution of the post-merger remnant \citep{Rosswog99,Bauswein13,Hotokezaka13}. This result was confirmed by the observations of the kilonova/macronova (hearafter kilonova) AT2017gfo associated to GW170817 \citep[e.g.][]{Andreoni2017, Coulter2017, Evans2017, Drout2017, Pian2017, Smartt2017, Tanvir2017, Troja2017, Valenti17}, 
showing that a large mass of ejecta was released by the NS merger. These ejecta expand at sub-relativistic velocities  and, at late times, are expected to
interact with the interstellar medium (ISM), producing a long lasting and nearly isotropic non-thermal radiation, peaking in the radio band \citep{np11,Piran+2013,hp15}. 
This signal could be detectable at radio wavelengths for events within $\approx$300 Mpc.
However, if the NS merger forms a massive and highly magnetized NS (magnetar) rather than a black hole (BH), the merger ejecta could be re-energized by the NS spin-power. This would result into a much brighter radio signal, visible at cosmological distances \citep{mb14}. Recent simulations of binary NS mergers indeed show that a stable magnetar can be formed for some ranges of initial NS masses \citep{gp13}. 
Such long-lived central engine is also invoked to model the X-ray afterglow of some short GRBs \citep{Fan2006,Rowlinson2010,Lu2015,Ciolfi2015} as well as  GW170817 \citep{Piro19}, and it might be a common outcome of NS mergers \citep{Piro2017}.

Long-term radio monitoring of short GRBs could be a powerful probe into the merger dynamics and remnant, complementary to kilonova and afterglow studies. The most promising target to detect the late-time radio (and possibly X-ray) counterpart of a NS merger is GW170817/GRB170817A, located at a distance of only 40 Mpc \citep{Hjorth2017}. This event is however relatively recent, and it may take years before detecting the onset of its remnant emission. 
Other local sGRBs, analogues to GRB170817A, could have been already discovered in past surveys but were likely not identified as nearby events due to the lack of precise (a few arcsec) afterglow localization.   \citet{Dichiara2020} searched for these events by cross-matching the list of sGRBs with no X-ray counterpart with the 
Galaxy List for the Advanced Detector Era \citep[GLADE,][]{Dalya2018}.
They selected a sample of four bursts, discovered by the \textit{Neil Gehrels Swift} Observatory \citep{Gehrels2004} that are  possibly located between 100 and 200 Mpc. 
No kilonova emission was detected in the optical for any of these bursts, implying either low ejecta masses of lanthanide-poor material or large viewing angles \citep{Dichiara2020}. No meaningful measurements were available in the infrared, thus leaving the mass of lanthanide-rich material unconstrained. If truly associated with local galaxies, as discussed in \citet{Dichiara2020}, these events would be prime targets to search for a late-time radio remnant.

All other known short GRBs lie instead at cosmological distances ($z\gtrsim$0.1), and their remnant could be detected only if the radio signal is boosted by the central magnetar. 
Past radio surveys of short GRBs started to place constraints to this model. 
\citet{mb14} presented a sample of seven bursts discovered between 2005 and 2007, and re-observed in 2008 with the Very Large Array (VLA). Due to the limited sensitivity of their observations ($\approx$0.3~mJy) and the short delay time between the burst and the observations, only mild constraints were derived. \citet{mb14} concluded that their limits were still consistent with a magnetar-powered radio flare for density values $n\lesssim$0.1~cm$^{-3}$, typical for the environment of sGRBs \citep{Oconnor2020}. 
Later observations, carried out with the Karl G. Jansky VLA \citep{Horesh2016,Fong2016, Schroeder2020} and the Australian Telescope Compact Array (ATCA; \citealt{Klose2019}), placed more stringent constraints for a larger sample of events. By assuming an ejecta mass $M\lesssim$0.01$M_{\odot}$, 
these new limits push the range of allowed densities to lower values ($n\lesssim$0.001~cm$^{-3}$) for an indefinitely stable magnetar with rotational energy $E\approx$10$^{53}$\,erg, whereas models with a lower energy reservoir or larger ejecta mass remain consistent with the observations for $n\lesssim$0.1~cm$^{-3}$ \citep[e.g.][]{Liu20}. These models peak at much later times, and could be better constrained by continued monitoring of sGRBs.

In this work, we present the results of a comprehensive radio monitoring of sGRBs, carried out with ATCA and the Jansky VLA, and including both cosmological and candidate local events. Selection criteria for the two samples are described in section~\ref{sec:sample}, observations and analysis in section \ref{sec:obs}, constraints on the circumburst density in section~\ref{sec:model}, light curve theory in section~\ref{sec:synchrotron}, constraints on ejecta in section~\ref{sec:ejecta} and our conclusions are drawn in section \ref{sec:conclusion}. 
Throughout this paper we adopt a standard $\Lambda$CDM cosmology \citep{Planck18}. Uncertainties are reported at the 68\% confidence level, 
upper limits at the 3~$\sigma$ level.

\section{Sample selection}\label{sec:sample}
\subsection{Cosmological sample}

We include in our study GRBs observed by {\it Swift} and classified as short bursts, based on their duration ($T_{90}$\,$\lesssim$2\,s) and spectral properties, 
as well as short bursts with extended emission \citep{nb06}.
We also include two long duration bursts, GRB~060614 and GRB~060505, whose classification is debated. In both cases, no bright supernova followed the GRB, instead a possible kilonova component was identified \citep{Ofek2007,Yang2015} suggesting that they were both produced by compact binary mergers. 

The sample of bursts and their properties are listed in Table~\ref{tab:sample}. 
All the GRBs in this sample have an arcsecond or sub-arcsecond localization. Their distance scale is estimated through their putative host galaxy, and is in all cases $z$\,$\lesssim$0.5.

Evidence for kilonova emission was discussed for
GRB~130603B, GRB~150101B, and GRB~160821B \citep{Tanvir2013,Troja2018,Troja2019,Lamb2019} as well as for GRB~060614 and GRB~060505 \citep{Ofek2007,Yang2015}, and is indicative of substantial mass ejection in these systems.  
For GRB~080905A, GRB~061201 and GRB~050509B, optical limits suggest instead that any kilonova emission would be much fainter than AT2017gfo \citep{Gompertz2018,Troja2018, Rossi20}, either because of a lower mass or a higher opacity of the merger ejecta. 
The distance scale of GRB~061201 is an additional factor of uncertainty in these estimates. 
In the remaining cases, no meaningful constraints could be derived either because the emission was dominated by the bright afterglow or because optical/nIR observations were not sufficiently sensitive.

For the majority of bursts in our sample ($\approx$60\%), a magnetar central engine was invoked to explain a wide variety of features: the temporally extended emission \citep{Gompertz2013}, the early steep decline (ESD) in
GRB080905A and GRB160821B \citep{Rowlinson2010,Lu2017,Troja2019}, the X-ray plateau of GRB~140903A \citep{Troja2016, LASKY}, and the long-lasting X-ray emission of GRB~130603B \citep{Fong2014}.

For most GRBs in our sample, rapid follow-up observations exclude the presence of an underlying radio source at the burst position \citep{Fong2015}. The probability of a radio transient occurring by chance at the same position and during our short observations would be negligible \citep{BowerSaul2011,Frail2012}.

\subsection{Local sample}

We use the sample of \citet{Dichiara2020}, consisting of four sGRBs:
GRB050906, GRB070810B, GRB080121, and GRB100206A. 
None of these events has an optical or X-ray counterpart, and therefore their best localization is the position from the \textit{Swift} Burst Alert Telescope (BAT), with a typical error radius of $\approx$2-3 arcmin. 
Within this error region, at least a nearby ($\lesssim$200~Mpc) galaxy was identified by \citet{Dichiara2020}: 
IC328 at $\approx$130~Mpc for GRB~050906, 
2MASX J00355339+0849273 at $\approx$175~Mpc for GRB~070810B, 
SDSS J090858.15+414926.5 and SDSS J090904.12+415033.2 at $\approx$200~Mpc for GRB~080121,
and LEDA 86918 at $\approx$172~Mpc for GRB100216A. 
Based on the GLADE catalogue, \citet{Dichiara2020} estimated that the chance alignment between nearby galaxies and BAT positions is $\approx$3\%, and the probability of finding four matches within their small sample is rather low. It seems therefore plausible that one or more of these bursts could belong to the local population of events. 

Radio archives were queried to check for the presence of any steady radio sources in the observing fields. Only one radio galaxy in the field of GRB 050906 is present in the the NVSS Catalogue \citep{Condon1998} as a weak (8.4 mJy) radio source at RA(J2000)=03:31:10.95, Dec(J2000)=-14:38:16.5.

\section{Observations and data reduction}\label{sec:obs}

\subsection{ATCA}
Observations with ATCA were carried out under projects C3059 and C3264 (PI: Troja) in the 16-cm band, with a center frequency of 2.1 GHz and band width of 2 GHz. The targets were observed between November 2015 and October 2018 using different array configurations. 
The observations were interleaved between target sources and suitably chosen phase calibrator every 20 min. The primary calibrator 1934$-$638 was used to calibrate the band-pass and bootstrap the absolute flux density scale. 
Details of the observing sessions and phase calibrators are provided in Table~\ref{tab:atcainfo}. 

The data were read into Miriad \citep{Sault1995} and split into single-source single-band data sets. These data sets were flagged for Radio Frequency Interferences (RFIs) and shadowing effects, calibrated and imaged using standard procedures in Miriad. 
The cleaned and restored maps of the targets were searched for radio transients at the GRB positions using the Karma software tool {\it kvis}. The typical rms noise in the I-Stokes restored maps was evaluated in {\it kvis} using rectangular regions away from bright sources. 

No significant radio emission was detected for any of our targets, except for a source of marginal significance ($<$5~$\sigma$) at the location of GRB~080123. 
This GRB field is however crowded with contaminating side sources, and a second observation of the same field did not confirm the tentative detection. 
In the latter observation the array configuration was much more suitable to carry out a search for faint transients (6A in Oct 2018 against EW352 in Jan 2016), and we conclude that the marginal signal seen during the first epoch was likely spurious. Results are listed in Table~\ref{tab:sample}.


\begin{table}
	\centering
	\caption{Log of ATCA observations: col. 1 source name, col. 2 observing date, col. 3 phase
        calibrator name, col. 4 array configuration, col. 5 observing notes: all bslns (all baselines used in in imaging), ant(6) only (only baselines with antenna 6 used in imaging).
       }
	\label{tab:atcainfo}
	\begin{tabular}{lllcc} 
		\hline
		Name     &       Date  &    Ph Cal              &     Array   &     Notes  \\
		\hline
		        GRB~130822A &     2015-10-30 &  0132-097     &         6A        &      all bslns  \\  
                GRB~150101B &     2015-10-30 &  1243-072     &         6A        &      all bslns  \\
                GRB~061201  &     2015-10-30 &  2142-758     &         6A        &      all bslns  \\
                GRB~100206  &     2015-11-02 &  0320+053     &         6A        &      all bslns  \\
                GRB~060614  &     2015-11-11 &  2052-474     &         6A        &      all bslns  \\
                GRB~060505  &     2015-11-11 &  2203-188     &         6A        &      all bslns  \\
                GRB~140903A &     2016-01-18 &  1607+268     &         750C      &      ant(6) only \\
                GRB~080905  &     2016-01-18 &  1908-201     &         750C      &      ant(6) only \\
                GRB~130603B &     2016-01-18 &  1117+146     &         750C      &      ant(6) only \\
                GRB~061006  &     2016-01-18 &  0606-795     &         750C      &      ant(6) only \\
                GRB 080123A &     2016-01-20 &  2117-642      &         EW352     &      ant(6) only \\
                           &     2018-10-23 &  2029-6910     &         6A        &      all bslns   \\          
		\hline
	\end{tabular}
\end{table}


\begin{table}
	\centering
	\caption{Log of VLA observations: col. 1 source name, col. 2 observing date, col. 3 phase calibrator name, col. 4 array configuration.}
	\label{tab:vlainfo}
	\begin{tabular}{lllcc} 
		\hline
		Name            & Date & Ph Cal  & Array \\
		\hline
		GRB050709   & 2016-05-06 &  J2214-3835 &  CnB   \\
		GRB~080905A  & 2016-05-03 & J1911-2006 &  CnB   \\
		& 2017-05-30 & J1911-2006 &  C         \\
		GRB150101B   & 2018-01-04 &  J1246-0730   &  B  \\
		GRB~100206A     & 2019-01-04 & J0309+1029 & C                   \\  
        GRB~130822A     & 2019-01-06 & J0125-0005 & C                    \\
        GRB~160821B     & 2019-01-08 & J1927+6117 & C                    \\
        GRB~050509B     & 2019-01-10 & J1221+2813 & C                    \\
        GRB~060502B     & 2019-01-11 & J1740+5211 & C                \\
        GRB~130603B     & 2019-01-12 & J1120+1420 & C                    \\
        GRB~150120A     & 2019-01-14 & J0029+3456 & C                    \\
        GRB~150424A     & 2019-02-05 & J1037-2934 & CnB                \\
        GRB~140903A     & 2019-02-05 & J1609+2641 & CnB                \\
        GRB~070810B     & 2019-08-07 & J0022+0608 &  A                   \\
        GRB~080121      & 2019-08-09 & J0920+4441 &  A                   \\        
        GRB~100216A     & 2019-08-09 & J1018+3542 &  A                   \\
        GRB~050906      & 2019-08-12 & J0340-2119 &  A                   \\
        GRB~170817A & 2019-08-11 & J1258-2219 & A \\
                    & 2019-08-30 & J1258-2219 & A \\
		\hline
	\end{tabular}
\end{table}

    
\begin{table*}
	\centering
	\caption{Observations of cosmological sGRBs: 
     col. 2 - redshift; 
     col. 3 - elapsed time since the GRB; 
     col. 4-5 - instrument and center frequency used; 
     col. 6 - upper limits (3$\sigma$ confidence level); 
     col. 7 - circumburst density; 
     col. 8 - presence of a candidate kilonova; 
     col. 9 - peculiar X-ray behavior (EE: Extended Emission; ESD: early steep decline); 
     col. 10 references. 
     Redshifts based on an uncertain identification of the GRB host galaxy are marked.}
	\label{tab:sample}
	\begin{tabular}{lccccccccl} 
	  \hline
	  Target name   &    $z$    & $T-T_0$ &  Array   &   $\nu$  &  Flux density     &  $n$      &   KN    &   X-rays &  References  \\
	                &           &   [d]   &           &   [GHz] &  [$\mu$Jy]        & [cm$^{-3}$] &         &           &       \\
          \hline
    GRB~050509B    &  0.225     & 4992    & VLA      &    6.0  &     < 6           &   $>2\times10^{-3}$      &         &          & \citet{Gehrels2005}  \\ 
    GRB~050709     & 0.160      &  3954   & VLA      & 3.0     & $<$31 & $<$ 0.1 & Y & EE & \citet{Jin16}\\
     GRB~060502B   &  0.287$^{\dag}$     & 4638    & VLA      &    6.0  &     < 6           &    $>8\times10^{-6}$     &         &          & \citet{Bloom2007}     \\ 
    GRB~060505     & 0.0894     &  3477   &  ATCA    &   2.1   &    < 111           &   $>4\times10^{-5}$  &   Y     &          & \citet{Ofek2007} \\
    GRB~060614     & 0.125      &  3437   &  ATCA    &   2.1   &     < 84           &  $>2\times10^{-4}$ &  Y     &    EE    & \citet{Yang2015} \\
    GRB~061006     & 0.436      &  3391   &  ATCA    &   2.1   &    < 183           &    $>0.01$       &         &    EE    & \citet{Davanzo2009} \\
    GRB~061201     & 0.111$^{\dag}$  &  3255   &  ATCA    &   2.1   &    < 117           &    $>0.2$      &         &          & \citet{Stratta2007}    \\
   GRB~080123      & 0.496      &  2919   &  ATCA    &   2.1   &      < 180       &    $>5\times10^{-3}$      &         &   EE     &             \citet{Klose2019}   \\    
                   &            &  3926   &  ATCA    &   2.1   &    < 123           &          &         &          &            \\
                   &            &  3926   &  ATCA    &   6.5   &     < 45           &        &          &          &           \\ 
  GRB~080905A      & 0.1218     &  2691    &  ATCA    &   2.1   &    < 279           &    $>7\times10^{-5}$     &          &  ESD & \citet{Rowlinson2010}  \\
                   &           &  2362    &  VLA     &   6.0   &     < 19           &           &          &          &            \\ 
                   &           &  2797    &  VLA     &   3.0   &     < 19           &         &          &          &            \\ 
                   &           &  3189    &  VLA     &   3.0   &     < 51           &         &          &          &            \\  
GRB 100206A         & 0.4068     &  2095    &  ATCA    &   2.1   &     < 48           &    $<16$     &          &    ESD      & \citet{Perley2012} \\
                   &           &  3254    &  VLA     &   6.0   &     < 12           &         &          &          &                \\ 
GRB 130603B         & 0.3565     &   959    &  ATCA    &   2.1   &    < 195           &   $5\times10^{-3}-30$     &   Y      &    late-time      & \citet{Tanvir2013}    \\
                   &           &  2139    &  VLA     &   6.0   &     < 11           &           &          &     excess     &     \\   
GRB 130822A     & 0.154$^{\dag}$&   799    &  ATCA    &   2.1   &     < 90           &    $<80$     &          &   ESD      & \citet{gcn15178}          \\
                   &           &  1963    &  VLA     &   6.0   &     < 9           &        &          &          &              \\
GRB 140903A         &  0.351    &   502    &  ATCA    &   2.1   &    < 153           & 0.006-0.17 &         & Plateau  & \citet{Troja2016} \\            
                   &           &  1617    &  VLA     &   6.0   &     < 11           &           &          &          &                \\ 
 GRB 150101B        & 0.1341    &   302    &  ATCA    &   2.1   &     < 78           & 0.001-0.6 &   Y      & Off-axis & \citet{Troja2018} \\
                    &           &  1109   & VLA      & 3.0 & 
                    $<$33 &       &          &        & \\
 GRB 150120A        & 0.460      &  1456    &  VLA     &   6.0   &     < 10           &    $<6$     &          &   ESD       & \citet{Chrimes2018} \\          
GRB 150424A         & 0.30$^{\dag}$       &  1384    &  VLA     &   6.0   &     < 15           &    $>0.01$     &          &   EE     & \citet{Jin2018}       \\  
GRB 160821B         & 0.1613     &   870    &  VLA     &   6.0   &     < 6           &    $5\times10^{-5}-0.02$     &     Y     & ESD  & \citet{Troja2019}     \\            
	  \hline
	\end{tabular}
\end{table*}


\subsection{VLA}

Observations with the VLA were carried out under projects 17A-248, 19A-194 (PI: Troja), 16A-159
(PI: Horesh), and 18B-168 (PI: Fong) in the C-band, with center frequency 6 GHz and nominal bandwidth of 4 GHz, and S-band, with center frequency 3 GHz and nominal bandwidth 2 GHz. 
The targets were observed between May 2016 and August 2019 using different array configurations. Details of these observing sessions are provided in Table~\ref{tab:vlainfo}.

Each target was observed for a total of one hour including set-up time, primary/band-pass calibrator scans, phase calibrator scans and target scans. 
Data were downloaded from the National Radio Astronomical Observatory (NRAO) online archive, and processed locally with the VLA automated calibration pipeline in CASA version 5.6.2. The calibrated visibilities for each target were further inspected and flagged to remove RFI, specially present in S-band. Finally, imaging was performed with the TCLEAN task, properly defining outliers cleaning windows when strong sources outside the imaged field were present. 
The CASA viewer task was used to evaluate RMS noise in regions of the restored maps away from bright radio sources. At the location of the sGRBs, no detections were found and 3\,$\sigma$ upper limits are provided in Table~\ref{tab:sample}.

\begin{figure*}
    \centering
    \begin{tabular}{cc}
  \includegraphics[width=80mm]{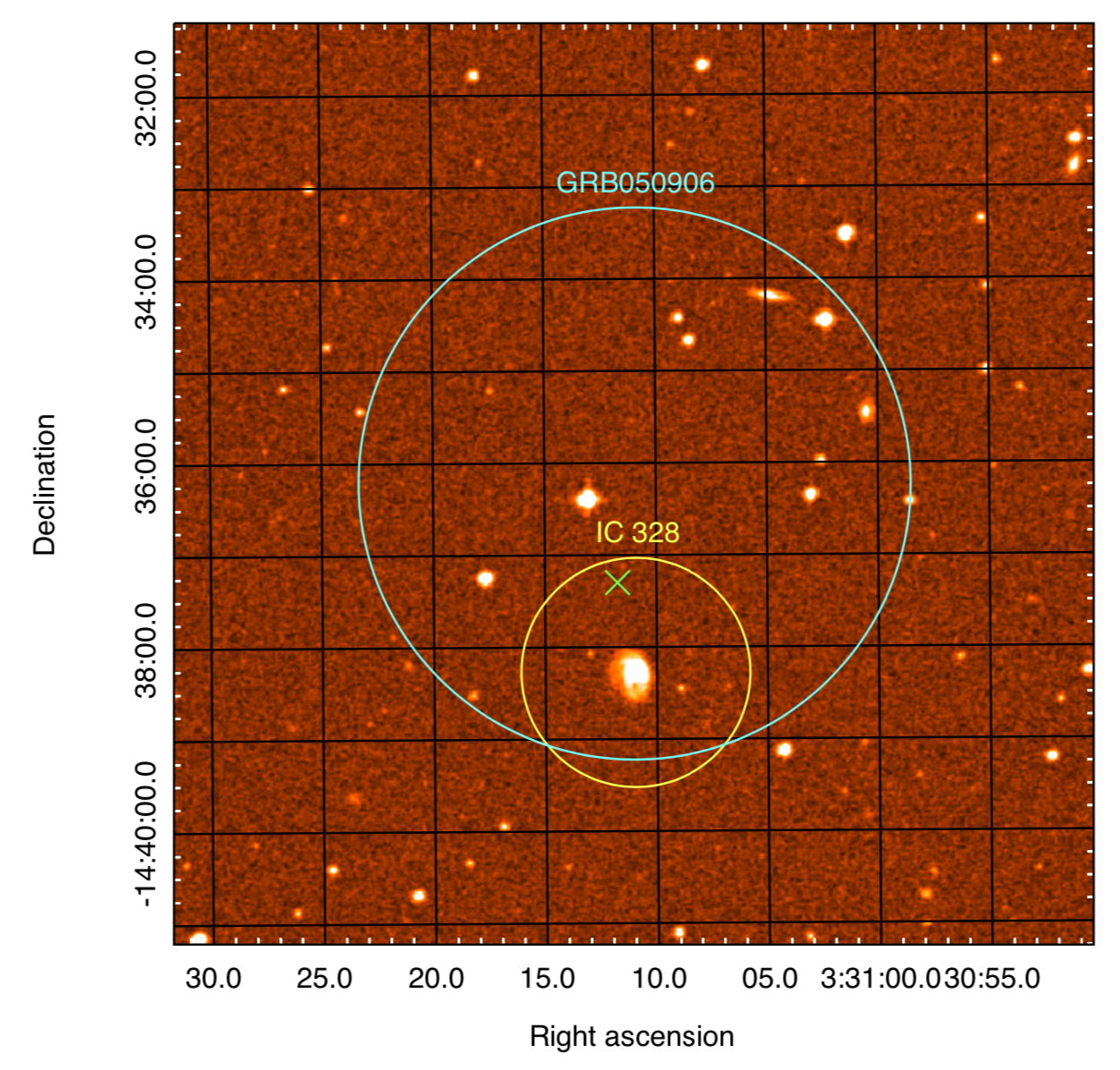} &   \includegraphics[width=79.3mm]{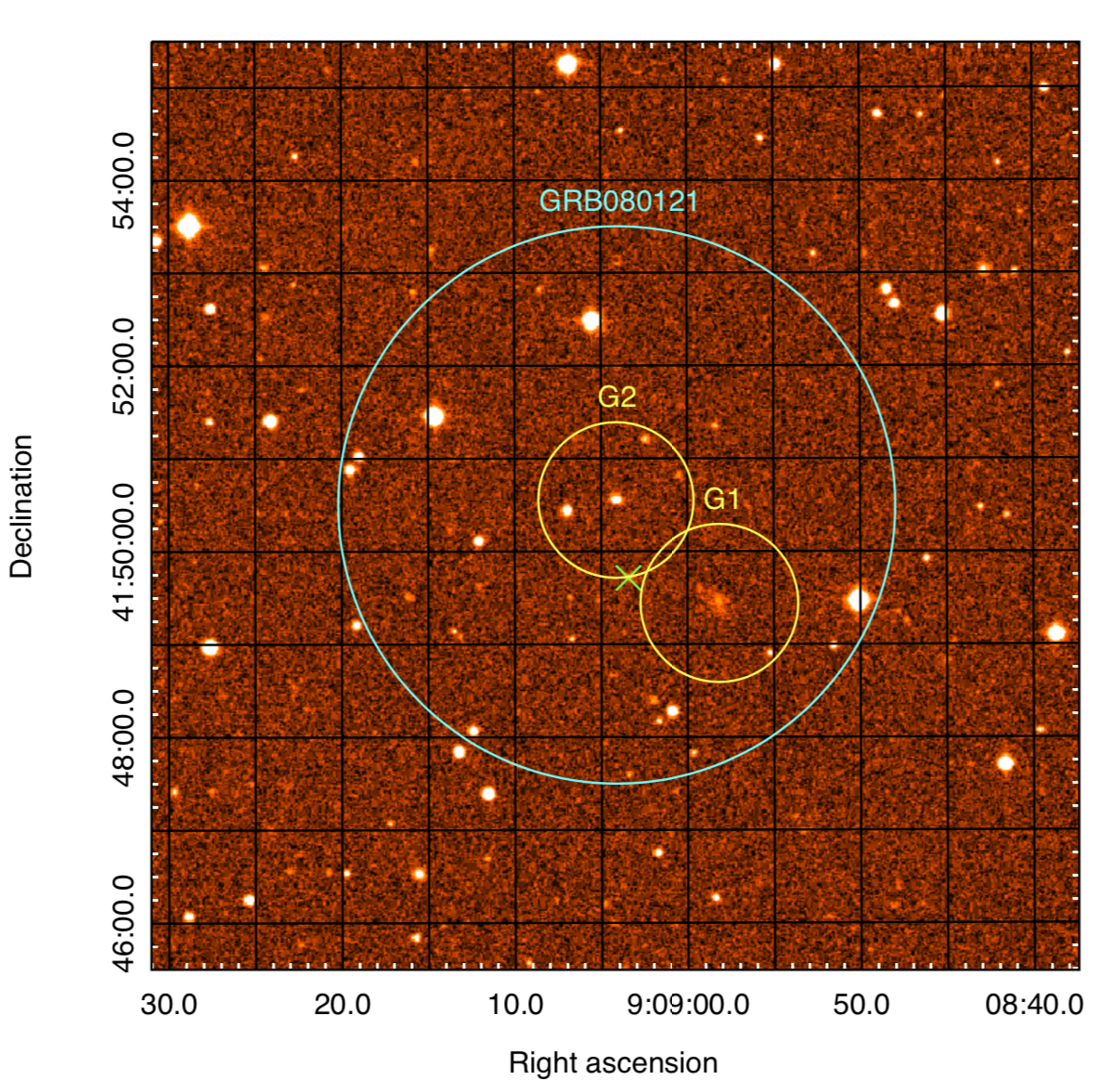} \\
\end{tabular}
    \caption{Optical fields for the two GRB from the local sample having a radio flare candidate: GRB050906 (left panel) and GRB080121 (right panel). The image is a 10$\times$10 arcmin DSS2-red field, with superimposed GRB position error diameter (6 arcmin, cyan), radio flare candidate position from our VLA observations (green cross), and a 50 kpc radius region (yellow) centered on the candidate host galaxies.}
    \label{fig:candidates}
\end{figure*}

For the sample of candidate local sGRBs, the VLA images were searched for any possible radio counterpart using the position of the putative host galaxy, and considering a maximum offset radius of 50~kpc. 
Results are reported in Table~\ref{tab:vlasample}: no radio source was found in association with GRB~070810B and GRB~100216A, 
whereas a candidate counterpart was found for GRB080121 and GRB050906 (Figure~\ref{fig:candidates}).
Sky radio source count surveys in \citet{Vernstrom2016} and \citet{Condon1984} were used to find the number of the expected sources in the 50 kpc region of the nearby galaxies: the chance coincidence probability is $\approx$0.6, consistent with the detection of two radio sources in a sample of four events. 
As an additional test, we compared the radio coordinates of these two candidates to optical catalogues.  
The candidate associated to GRB~080121 lies at the center of a bright extended source visible in Pan-STARRS images \citep{panstarrs}, and thus is most likely due to AGN activity. 
No optical match was found at the position of the radio candidate in the field of GRB~050906. However, the observed radio spectrum of this source does not match our expectations of optically thin synchrotron emission. Therefore, also in this case, we find no evidence supporting a physical connection with the GRB or the scenario of a late-time radio flare. 

For completeness, we also included in our analysis recent VLA observations of GRB170817A/GW170817 taken in August 11th and 30th, 2019 (Project SK0299, PI: Margutti) at 6 GHz with a nominal bandwidth of 4 GHz. 
Data were downloaded from the archive after being calibrated via the CASA online calibration pipeline v5.4.0. The two calibrated measurement 
sets was then split, concatenated using task {\it concat} and imaged with the task {\it clean} using a Briggs parameter value of 0.5 and 5000 iterations. 
In agreement with the result reported by \citet{Hajela2019}, no detection was found at the target position and a 3\,$\sigma$ flux density upper limit of 6.9~$\mu$Jy was derived using the task {\it imstat}
in a region around the target position.

\begin{table*}
	\centering
	\caption{Local sample results. Col. 2: elapsed time in the observer's frame; col.3-4: 3-$\sigma$ upper limits measured at S-band (3~GHz) and C-band (6~GHz); col.5: position of the candidate radio counterpart; col. 6-7: integrated flux density of the radio candidates at 3 and 6~GHz (in case of non detection an upper limit is provided at the location of the radio candidate); col. 8: galaxies within the GRB localization; col. 9: galaxy's distance.}
	\label{tab:vlasample}
	\begin{tabular}{lccllccclc} 
	  \hline
    GRB name  &    t-t$_0$    &    3 GHz UL  & 6 GHz UL & 
    Candidate Location &  Flux(3GHz) & Flux(6GHz)   & 
    Nearby Galaxy  & Distance \\
            &    [d]        &    [$\mu$Jy]     & [$\mu$Jy]    & 
    [RA/DEC J200]            & [$\mu$Jy]       & [$\mu$Jy]       
    &         &  [Mpc]        \\
     \hline
    GRB 070810B      &    4380       &    < 20      & < 14     &   $--$             &   $--$      &   $--$       &    2MASX J00355339+0849273     & 175   \\
    GRB 100216A       &    3461       &    < 20      & < 18     &   $--$             &   $--$      &   $--$       &    LEDA 86918    & 172   \\
    GRB 050906       &    5088       &    < 53      & < 14     &  03:31:11.746      &  < 272      & 640 $\pm$ 32 &  IC 328   & 130  \\
                         &               &              &          & -14:37:17.960      &             &              &     &        \\    
    GRB 080121    &    4218       &    < 26      & < 12     & 09:09:03.431       & 130 $\pm$ 11 & 110 $\pm$ 8 & SDSS J090858.15+414926.5 (G1) & 200 \\
                   &               &              &          & 41:49:42.786       &              &            & SDSS J090904.12+415033.2  (G2) & 200 \\
          \hline
	\end{tabular}
\end{table*}


\section{Constraints on Circumburst Density}\label{sec:model}

The properties of the late-time radio emission depend on the details of the explosion, such as the total mass of ejecta $M_\textrm{ej}$ and their velocity profile, the shock parameters, and the density $n$ of the surrounding medium. The latter value can be independently constrained by using the observed afterglow emission, arising from the interaction of the GRB jet with the circumburst environment. 

When available, we used the estimates from broadband afterglow modeling (GRB~050709, \citealt{Panaitescu06}; GRB130603B, \citealt{Fong2014}; GRB140903A, \citealt{Troja2016}; GRB150101B, \citealt{Troja2018}; GRB~160821B, \citealt{Troja2019}). However, the majority of sGRBs have weak afterglows, detected only in the X-ray band. In these cases, \citet{Oconnor2020} used the early X-ray light curves to constrain the afterglow peak, and set a lower limit, $n_\textrm{min}$, to the density. The peak of the afterglow light curve occurs when the blast-wave has encountered enough ambient medium that the initial bulk Lorentz factor of the jet begins to decrease significantly \citep{Blandford1976}. Therefore, the peak of the afterglow contains information on the density of the surrounding medium. 
Following \citet{Oconnor2020}, we combine the constraints on the peak time and the peak flux
to derive the minimum allowed density, $n_\textrm{min}$. This method applies to GRB afterglows consistent with forward shock emission, but cannot be used in other cases when the observed light
is dominated by different emission components, such as early steep declines or internal plateaus. 

The standard afterglow model \citep{Sari1998}, 
in which electrons accelerated in the forward shock emit synchrotron radiation, is described by a set of six parameters:
\{$n$, $\widetilde{p}$, $\widetilde{\varepsilon}_B$, $\widetilde{\varepsilon}_e$, $\widetilde{E}_\textrm{k,iso}$, $\widetilde{\Gamma}$\}, where $n$ is the circumburst 
particle density, $\widetilde{p}$ is the slope of the emitting electrons' power-law energy distribution $N(\gamma)\propto \gamma^{-\widetilde{p}}$, $\widetilde{\varepsilon}_B$ and $\widetilde{\varepsilon}_e$ are the fractions of the burst kinetic energy $\widetilde{E}_\textrm{k,iso}$ that exist in the magnetic field and electrons respectively, and $\widetilde{\Gamma}$ is the initial bulk Lorentz factor of the jet. 
The tilde symbol is used to denote the GRB afterglow parameters, 
and distinguish them from those describing the ejecta emission.

In our calculations, we adopt $\widetilde{p}=2.2$ and $\widetilde{\varepsilon}_e=0.1$, which has a narrow distribution ($\sigma_{\log\widetilde{\varepsilon}_e}\sim0.3$) centered around this value \citep{Nava2014,BeniaminiVanDerHorst}. 
In order to convert the observed gamma-ray energy into the blast-wave's kinetic energy, $\widetilde{E}_\textrm{k,iso}$, we adopt a gamma-ray efficiency of $\eta_\gamma=0.15$, similarly shown to have a narrow distribution \citep{Nava2014,Beniamini2015}. This is a conservative approach as higher values of efficiency would result in tighter lower limits. The fraction of energy in the magnetic fields, $\widetilde{\varepsilon}_B$, is less constrained. We apply $\widetilde{\varepsilon}_B=$0.01 as a conservative solution,  decreasing $\widetilde{\varepsilon}_B$ would only increase the minimum density allowed. 

The bulk Lorentz factor is also poorly constrained due to the difficulty inferring its value from afterglow modelling. A value of $\widetilde{\Gamma}$\,$\approx$1,000 was derived for the short GRB 090510 \citep{LAT090510}, however bursts with bright GeV emission tend to sample the most relativistic explosions and they are not representative of the general population of short GRBs. 
More typical values derived for long GRBs are of order a few hundred \citep{Ghirlanda2018} and we adopt $\widetilde{\Gamma}\approx 100$ for sGRBs considered in this work. The lower limits are presented in Table \ref{tab:sample}. 

In three events (GRBs 100206A, 130822A, and 150120A) the X-ray afterglow is only detected during the initial phase of steep decline, 
usually associated to long-lived central engine activity \citep[e.g.,][]{Nousek2006}.
Thus any contribution from the forward external shocks should be below the observed X-ray flux. We estimate upper limits on circumburst density $n$  by assuming that the accelerated electrons, emitting synchrotron radiation in the X-ray band ($0.3-10$ keV), are in the slow cooling regime (i.e., $\nu_X<\nu_c$, where $\nu_c$ is the synchrotron cooling frequency). We consider this a valid assumption for typical sGRB parameters \citep[see, e.g., Figure 2 in][]{Oconnor2020}.

The resulting upper limits, also reported in Table \ref{tab:sample}, rule out only the highest density environments. This is due to the degeneracy between $n$ and $\widetilde{\varepsilon}_B$ ($n\propto\widetilde{\varepsilon}_B~^{-8/5}$ for $p=2.2$), which is a poorly constrained parameter. In our calculations we conservatively adopt 
$\widetilde{\varepsilon}_B$\,$\approx$10$^{-5}$, which is the lowest value inferred for a sGRB afterglow. Higher values of $\widetilde{\varepsilon}_B$ would bias the result to lower densities.  For example, for the short GRB~150120A,
we derive that an upper limit $n$\,$\lesssim$6~cm$^{-3}$ is consistent with the afterglow data, whereas
a value of  $\varepsilon_B\approx$0.01-0.1, commonly used in the literature \citep[e.g.,][]{Fong2015, Schroeder2020},  would only allow for low-density solutions, $n$\,$\lesssim$0.01-0.001 cm$^{-3}$.

\section{The Radio Light curve }\label{sec:synchrotron}

We turn now to estimate the radio emission from the merger's ejecta. This radio emission arises from the interaction of the ejecta with the surrounding circum-burst matter \citep{np11}. We assume for simplicity that the ejecta is expanding spherically with a single initial velocity, $\betain$. The ejecta is characterized therefore by its kinetic energy $\Eej$ and mass $\Mej$. While the latter does not appear explicitly in the calculation it determines the velocity through 
\begin{align}
\Eej\simeq\Mej c^2(\Gamin-1),
    \label{eq:gamma_in}
\end{align}
where $\Gamma_{\rm in}=1/\sqrt{1-\betain^2}$ is the Lorentz factor corresponding to $\betain$. Other parameters that determine the ejecta emission are the circumburst density $n$ and the microphysical parameters that determine the synchrotron emission from the shock wave: $\epe$ and $\epb$, the equipartition parameters of the electron's energy and magnetic field energy and $p$, the slope of the electron injection spectrum. 
It is important to stress that, while the shocks that produce the sGRB afterglow are relativistic, the shocks due to the interaction of the ejecta with the circumburst ISM are Newtonian. Hence the microphysical parameters controlling these two sets of shocks could be very different. 

In addition to the these parameters the resulting light curve could be quenched  by the blast  wave produced by the sGRB jet \citep{Margalit&Piran2020}. The effect of this blast wave is characterized by a single parameter $\Ej$: the overall energy of the jet. 

Our goal in this work is to constrain,  using the late time radio upper limits, the energy of the ejecta. As can be clearly understood from the above discussion the radio light curve depends on numerous parameters some of which are unknown. Among those parameters, the ISM density, ejecta mass, and the magnetic equipartition are most important. In the earlier section we have obtained constraints to the ISM density that we use here. 
As for the ejecta mass we have some estimates in cases that a macronova/kilonova candidate was observed \citep[see][for a compilation of mass estimates for such candidates]{Ascenzi+2019}. The last parameter, $\epb$ is the least determined. GRB and radio supernova observations bracket it in a wide range $10^{-5}$ to $10^{-1}$. We consider in the following only $\epb=0.1$ and $0.01$ as with lower values the radio signal would be too weak.

\subsection{Dynamics of ejecta}
The time evolution of the velocity of ejecta is determined by the energy conservation \citep{Piran+2013,hp15}
\begin{align}
\Eej&=[\Mej+\Gamma M(R)]c^2(\Gamma-1),
\end{align}
where $M(R)=4\pi \MP n R^3/3$ and $\MP$ are the mass of ISM swept-up by the ejecta at radius $R$ and the proton mass, respectively.
The radius of ejecta is obtained by integrating the velocity:
\begin{align}
R(t)&=\int_0^tdt^\prime\frac{\beta(t')  c}{1-\beta(t')},
\end{align}
where the denominator, $1-\beta$, is included to take the relativistic effect into account.
While the ejecta we consider are Newtonian or at most mildly relativistic, the above formalism enables us to calculate the dynamics of both relativistic and Newtonian ejecta. 

For the Newtonian case, we can obtain more simplified formulae for the ejecta radius and velocity.
The ejecta expand freely at an early time but start to decelerate when $M(R)=\Mej$. This timescale is given by
\begin{align}
\tdec&\simeq\biggl(\frac{3\Eej}{2\pi\MP n c^5\betain^5}\biggl)^{1/3}\simeq5.4{\,\rm yr\,}\Eejn^{1/3} n_0^{-1/3}\biggl(\frac{\betain}{0.3}\biggl)^{-5/3},
\end{align}
where we use the convention of $Q_x=Q/10^x$ (cgs).
Note that the radio light curve also peaks at this timescale.
Since the following expansion is well approximated by the Sedov-Taylor blast wave, we obtain 
\begin{align}
R(t)&\simeq\begin{cases}\betain c t&:t<\tdec,\\
\betain c \tdec\Big(\frac{t}{\tdec}\Big)^{2/5}&:\tdec<t,
\end{cases}
    \label{eq:radius}\\
\beta(t)&\simeq\begin{cases}
\betain&:t<\tdec,\\
\betain\Big(\frac{t}{\tdec}\Big)^{-3/5}&:\tdec<t.
\end{cases}
    \label{eq:velocity}
\end{align}

Had we ignored the jet dynamics these equations would have determined completely the radio light curve of the ejecta. However, 
following  \cite{Margalit&Piran2020}, we consider the interaction between the ejecta and the jet component. 
The jet dynamics is characterized by $\tj$:
\begin{align}
\tj&=\biggl(\frac{\Ej}{\MP nc^5}\biggl)^{1/3}\simeq0.2{\,\rm yr\,}\Ejn^{1/3}n_0^{-1/3}.
\end{align}
The jet becomes Newtonian at $\tnr$:
\begin{align}
\tnr&\sim\biggl(\frac{3\Ejiso}{4\pi \MP nc^5}\biggl)^{1/3}\simeq0.57{\,\rm yr\,}E_{\rm j,iso,51}^{1/3}n_0^{-1/3} \simeq3.6\,\theta_{\rm j,-1}^{-2/3} \tj.
\end{align}
At $t \ge \tnr$ \citep{Margalit&Piran2020} the jet becomes quasi-spherical producing  a Sedov-Taylor blast wave:
\begin{align}
R_{\rm j}(t)&=\xi c\tj\biggl(\frac{t}{\tj}\biggl)^{2/5},
	\label{eq jet r}\\
\beta_{\rm j}(t)&=\frac{2}{5}\frac{R(t)}{ct}=\frac{2\xi}{5}\biggl(\frac{t}{\tj}\biggl)^{-3/5},
	\label{eq jet beta}
	\end{align} 
where $\xi=1.17$.
This blast wave ahead of the merger ejecta sweeps up the ISM and quenches the ejecta radio signature. 
Thus we will not detect any emission from the ejecta until it collides with the jet at 
\begin{align}
\tcol&=\biggl(\frac{\xi}{\betain}\biggl)^{5/3}\tj \simeq1.9{\,\rm yr\,}\Ejn^{1/3}n_0^{-1/3}\biggl(\frac{\betain}{0.3}\biggl)^{-5/3}.
\end{align}

\subsection{The Synchrotron flux}
We turn now to calculate the resulting  synchrotron emission.
Given that we are considering only relatively late signals we ignore in the text (but not in the actual calculations when relevant) the possibility that the ejected mass is moving relativistically.
Similarly we ignore any redshift effects in the following equations but these are included in the actual numerical calculations.
The synchrotron flux is estimated by modification of the extreme relativistic equations of \cite{Sari1998}  \citep[see also][] {Chevalier1982,Chevalier1998,Piran+2013}. 
The magnetic field is estimated by the equipartition argument:
\begin{align}
B&=(8\pi\epb\MP nc^2\beta^2)^{1/2}
\simeq 0.018 {\,\rm G\,}\epbn^{1/2}n_0^{1/2}\biggl(\frac{\beta}{0.3}\biggl) \ . 
\end{align}
We assume that electrons are accelerated at the shock and an non-thermal power-law distribution of the electrons' Lorentz factor $\gamma$ is realized, $dN/d\gamma\propto\gamma^{-p}$.
The minimal electron's Lorentz factor, $\gamm$, is also determined by the equipartition argument:
\begin{align}
\gamm&=\frac{\MP}{\ME}\biggl(\frac{p-2}{p-1}\biggl)\epe\beta^2\simeq4.1\,\bepen\biggl(\frac{\beta}{0.3}\biggl)^2,
    \label{eq:gamma_m}
\end{align}
where $\ME$ is the electron mass and we have 
redefined the equipartition parameter as $\bepe\equiv4\epe(p-2)/(p-1)$. The characteristic synchrotron frequency of an electron with $\gamm$ and the corresponding flux are given by
\begin{align}
\num&=\gamm^2\frac{eB}{2\pi\ME c}
\simeq8.7\times10^{5}{\,\rm Hz\,}\,\bepen^2\epbn^{1/2}n_0^{1/2}\biggl(\frac{\beta}{0.3}\biggl)^5,
    \label{eq:nu_m} 
    \\
F_{\num}&=\frac{R^3n}
{3 d^2}\frac{\SGM c \gamm^2 B^2}{6\pi \num}
\simeq0.23{\,\rm Jy\,}\epbn^{1/2}n_0^{3/2}d_{27}^{-2}R_{18}^3\biggl(\frac{\beta}{0.3}\biggl),
    \label{eq:f_m}
\end{align}
where $e$ and $\SGM$ are the elementary charge and the Thomson cross section, respectively and $d$ is the distance to the burst.
The synchrotron self absorption frequency is given by   \citep{Rybicki&Lightman1979,Murase+2014}:
\begin{align}
\nua&=\biggl[\frac{(p-1)\pi^{\frac{3}{2}}3^{\frac{p+1}{2}}}{4}\frac{enR}{B\gamm^5}\biggl]^{\frac{2}{p+4}}\num \nonumber \\
&\simeq3.6\times10^{8}{\,\rm Hz\,}\bepen^{\frac{2(p-1)}{p+4}}\epbn^{\frac{p+2}{2(p+4)}}n_0^{\frac{p+6}{2(p+4)}}R_{18}^{\frac{2}{p+4}}\biggl(\frac{\beta}{0.3}\biggl)^{\frac{5p-2}{p+4}}.
    \label{eq:nu_a}
\end{align}
We find that the observed frequency becomes smaller than $\nua$ only for relativistic ejecta.
When calculating  the flux, we use the fact that $\nu>\num,\,\nua$ within the typical parameters we consider. Even though the observed frequency is much larger than both $\num,\,\nua$ it is much lower than the cooling frequency so it is within a single spectral slope and the flux is determined by $F_{\num}$ and $\nu/\num$. 

Within most of the parameter ranges we consider, we have $\num<\nua$ and the synchrotron spectrum is given by \citep[e.g.][]{Piran+2013} 

\begin{align}
F_{\nu}=\begin{cases}
F_{\num}(\nua/\num)^{\frac{1-p}{2}}(\nu/\nua)^{5/2}(\nu/\num)^{2}&:\nu<\num,\\
F_{\num}(\nua/\num)^{\frac{1-p}{2}}(\nu/\nua)^{5/2}&:\num<\nu<\nua,\\
F_{\num}(\nu/\num)^{\frac{1-p}{2}}&:\nua<\nu.\\
\end{cases}
    \label{eq:spectrum}
\end{align}

\begin{figure}
\begin{center}
\includegraphics[trim = 5 10 0 0, clip,width=\columnwidth]{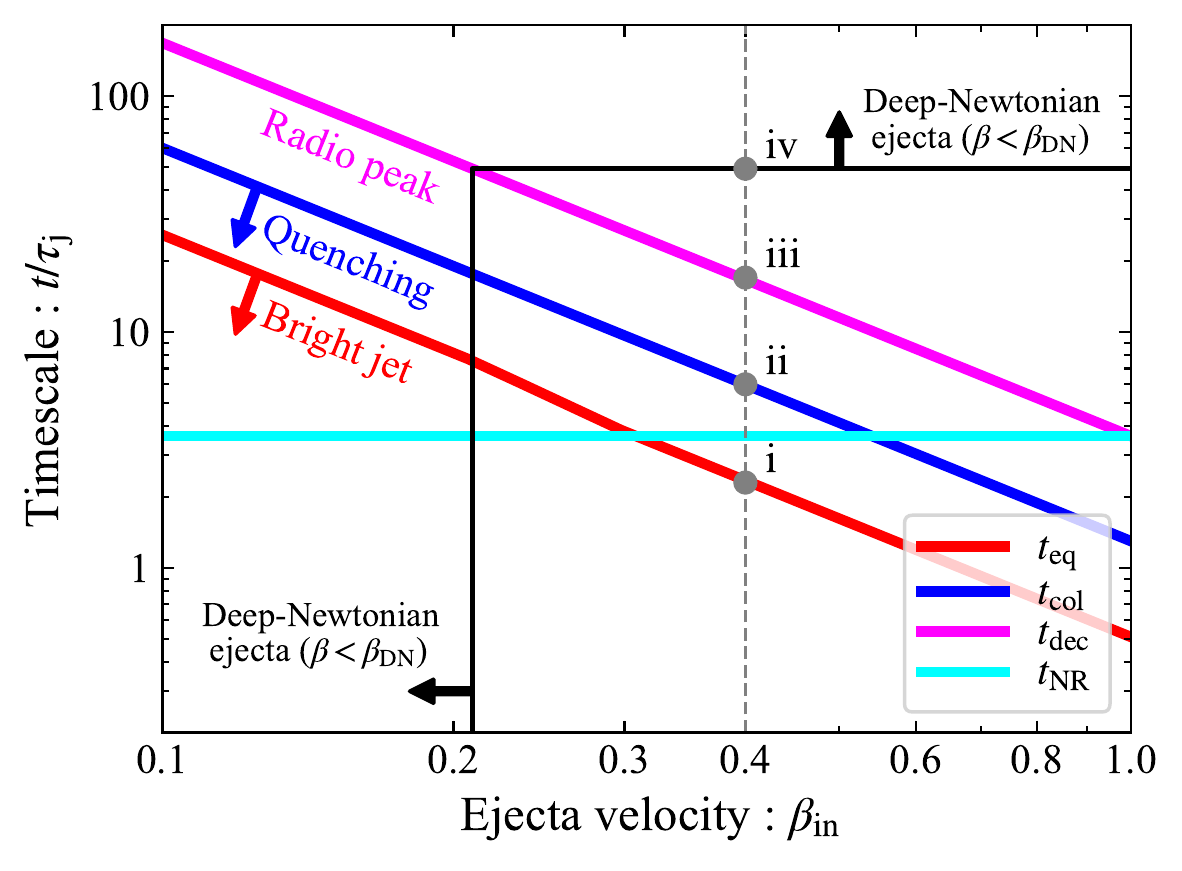}
\caption{ Normalized timescales as a function of $\betain$, the ejecta initial velocity for  $\Eej/\Ej=100$. 
The horizontal cyan line depicts $\tnr$, the  transition time of the jet to Newtonian regime.  The radio flare peaks on the magenta line and it declines (rises) above (below) this line. Below the blue line, the emission from the ejecta is quenched by the jet. On the red line, the (unperturbed) ejecta signature would have been equal to the jet afterglow and the latter dominates below this line.  
The dashed vertical line outlines the behaviour of the light curve for a given
initial velocity ($\betain=0.4$). At early time below (i) the afterglow radio signal dominates. Between (i) and (ii) the jet blast-wave that has  expanded to a quasi-spherical structure quenches the ejecta radio signal. The ejecta collides with the jet blast-wave at (ii) and from that time on its interaction with the ISM produces a radio light curve  peaks at (iii). It undergoes a transition to the deep-Newtonian regime at (iv). 
}
\label{fig time}
\end{center}
\end{figure}

When the Lorentz factor $\gamm<2$, the emission is dominated by electrons with $\gamma\sim2$.
Thus we can effectively fix $\gamm=2$ but the fraction of emitting electrons is reduced to $(\beta/\bdn)^2$, where
the critical velocity below which the deep-Newtonian phase begins, is given by setting $\gamm=2$ in Eq. \eqref{eq:gamma_m}:
\begin{align}
\bdn=\biggl(\frac{80\ME}{\MP}\biggl)^{1/2}\,\bepen^{-1/2}\simeq0.21\,\bepen^{-1/2}.
\end{align}
During this phase, Eqs. \eqref{eq:nu_m}, \eqref{eq:nu_a}, and \eqref{eq:f_m} are modified as
\begin{align}
\num&=\frac{ 2 eB}{\pi\ME c} \simeq6.9\times10^{4}{\,\rm Hz\,}\,\epbn^{1/2}n_0^{1/2}\beta_{-1},\\
\nua&=\biggl[\frac{(p-1)\pi^{\frac{3}{2}}3^{\frac{p+1}{2}}}{4}\frac{enR(\beta/\bdn)^2}{B\gamm^5}\biggl]^{\frac{2}{p+4}}\num\\
&\simeq3.9\times10^{7}{\,\rm Hz\,}\bepen^{\frac{2p}{p+4}}\epbn^{\frac{p+2}{2(p+4)}}n_0^{\frac{p+6}{2(p+4)}}R_{18}^{\frac{2}{p+4}}\beta_{-1}^{\frac{5p+2}{p+4}}, \nonumber \\
F_{\num}&=\frac{R^3n(\beta/\bdn)^2P_{\num}}{3 d_{\rm L}^2}\\
&\simeq 0.018 {\,\rm Jy\,}\bepen\epbn^{1/2}n_0^{3/2}d_{\rm L,27}^{-2}R_{18}^3\beta_{-1}^3, \nonumber 
\end{align}
respectively.

The resulting light curve is determined by three time scales (see Figure \ref{fig time}):
\begin{itemize}
\item{}  $\teq$ -  when  the rising ejecta light curve would have been equal to the
declining jet afterglow radio signal    (if the former was not quenched due to the jet, as mentioned above); 
\item{} $\tcol$ -  when the ejecta collides with the Sedov - Taylor blast wave produced by the jet and its radio signal emerges; 
\item{} $\tdec$  - when the ejecta radio signal peaks. 
\end{itemize}
It is convenient to normalize these  time scales by $\tj$. 
\begin{align}
\frac{\teq}{\tj}\simeq\begin{cases}
26\,\Big(\frac{\betain}{0.1}\Big)^{-5/3}&:\betain\lesssim\bdn,\\
5.4\,\bepen^{-\frac{10}{81}}\Big(\frac{\betain}{0.25}\Big)^{-\frac{35}{27}}&:\bdn\lesssim\betain\lesssim0.3\bepen^{-1/2},\\
1.6\biggl(\frac{\betain}{0.5}\biggl)^{-5/3}&:0.3\bepen^{-1/2}\lesssim\betain.\\
\end{cases}
\end{align}
The numerical factors and the powers derived here correspond to $p=2.5$  and $\xi=1.17$. We also assume that the microphysical parameters such as $\varepsilon_{\rm e}$ and $\varepsilon_{\rm B}$, are common for the ejecta and the Newtonian jet.   

The radio fluxes and the equality time depend on the corresponding velocities of the ejecta and the jet blast-wave. 
In the first case, both the blast-wave and ejecta are in the deep-Newtonian case.
In the second case, only the jet is in the deep-Newtonian case. In the last case, the fluxes become equal before both jet and ejecta are in the deep-Newtonian regime.
The two other time scales are: 
\begin{align}
\frac{\tdec}{\tj}&\simeq170\,\Ejn^{-1/3}\Eejn^{1/3}\beta_{\rm in,-1}^{-5/3},\\
\frac{\tcol}{\tj}&\simeq\xi^{5/3}\betain^{-5/3}\simeq60\,\beta_{\rm in,-1}^{-5/3}.
\end{align}

Finally, we note that for $\betain>\bdn$, the ejecta enters  the deep-Newtonian phase when 
\begin{align}
\frac{t_{\rm DN,ej}}{\tj}\simeq 50\,\bepen^{5/6}\Ejn^{-1/3}\Eejn^{1/3}.
\end{align}

Figure \ref{fig time} depicts each timescale as a function of the initial velocity $\betain$. It shows the different timescales of the possible light curves - along a vertical line, if $\betain$ is known -  or the possible interpretation of the observations - along a horizontal line, if the observations time is known.

In the following, in order to understand the results, it is useful to consider the light curve  before and  after the peak. 
For $\nu>\num,\,\nua$ and $\beta>\bdn$
the behaviour is given by:

\begin{align}
F_{\nu}\simeq\begin{cases}
27{\,\rm \mu Jy\,}\bepen^{p-1}\epbn^{\frac{p+1}{4}}n_0^{\frac{p+5}{4}}\Big(\frac{\betain}{0.3}\Big)^{\frac{5p+3}{2}}d_{27}^{-2}\nu_{\rm GHz}^{\frac{1-p}{2}}t_{\rm yr}^3&:t<\tdec,\\
6.7\times10^{4}{\,\rm \mu Jy\,}\bepen^{p-1}\epbn^{\frac{p+1}{4}}n_0^{\frac{19-5p}{20}}\Eejn^{\frac{5p+3}{10}}d_{27}^{-2}\nu_{\rm GHz}^{\frac{1-p}{2}}t_{\rm yr}^{\frac{3(7-5p)}{10}}&:\tdec<t,
\end{cases}
\label{eq:flux}
\end{align}
where $t=t_{\rm yr}\,\rm yr$ and $\nu=\nu_{\rm GHz}\,\rm GHz$.
Notice a roughly linear dependence on $\bepe$. The dependence on $n$ is almost quadratic before the peak but then it is much weaker after it (see also Figures \ref{fig:lc} and \ref{fig:lumi}).

\section{Light curves and constraints} \label{sec:ejecta}

We use the above methodology to calculate the ejecta's radio light curves for the different mergers. 
Comparing these light curves to  the late time radio observations we constrain the parameters of the ejecta  and in particular its energy. When calculating light curves we must set some parameters to values for which we can carry out the comparison. In the following we choose the following as fiducial values:
\begin{itemize}
\item{}  the external density, $n$ is set for each event to different values according to the limits shown in Table \ref{tab:sample}.
\item{}the electron's equipartition fraction is $\bepe=0.1$.
\item{} the magnetic equipartition fraction $\epb$, one of the least constrained parameters is set to 0.1 and at times we consider also 0.01.
\item{}the power-law index of the electrons' distribution is $p=2.5$.
\item{}we consider three values for the ejecta energy $\Eej=10^{51}, 10^{52}$ and $10^{53}$ erg.
\end{itemize}
We do not set the ejected mass to a specific value. Instead we consider different ejecta velocities, which are related with $\Eej$ and $\Mej$ through Eq. \eqref{eq:gamma_in}. Finally, when there is no specific information about a given burst, we choose a jet energy $\Ej=10^{49}$ erg. 

\begin{figure}
\begin{center}
\includegraphics[width=85mm, trim = 5 2 5 2, clip]{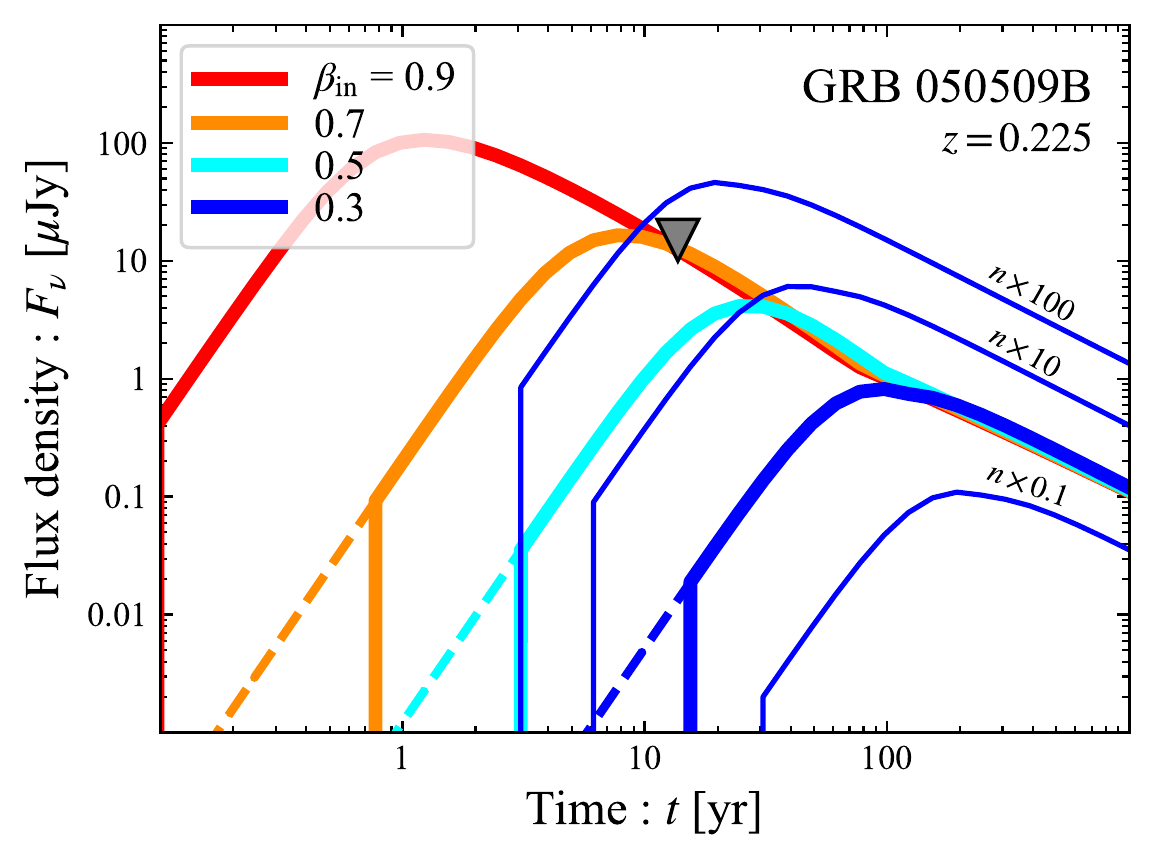}
\vspace{-0.5cm}
\caption{Light curve of GRB~050509B for various initial velocities $\betain$, 
an energy $\Eej=10^{52}$ erg and ISM density $n=2 \times 10^{-3} {\rm cm}^{-3}$, the lowest value allowed from afterglow data (see Table \ref{tab:sample}). 
Other parameters are: $p=2.5$, $\bepe=0.1$, and $\epb=0.1$.
The gray downward triangle shows the observed radio upper limit.
The vertical branch describes the quenching due to the jet blast-wave. The dashed branch describes the light curve if this quenching did not take place. For the case $\betain$=0.3 (blue), the narrow solid curves show how the lightcurve changes for different densities.
Note that with this energy, $\Mej=0.03\Msun$ corresponds to $\betain\simeq0.5$.}
\label{fig:lc}
\end{center}
\end{figure}

We begin with an example of GRB~050509B \citep{Gehrels2005}.
Figure \ref{fig:lc} depicts light curves from the ejecta of GRB 050509B for various initial ejecta velocities. For  the chosen ejecta energy, $\Eej=10^{52}$ erg, the velocities, $\betain=0.9$, 0.7, 0.5, and 0.3, correspond to ejecta masses of $\Mej\simeq4.3\times10^{-3}$, $1.4\times10^{-2}$, $3.6\times10^{-2}$ and $1.2\times10^{-1}\,\Msun$, respectively. 
Ejecta with velocities $\lesssim$0.3 would produce a weak and delayed radio transient, not probed by our observations. 
The density $n=2 \times 10^{-3} {\rm cm}^{-3}$ is the minimal one estimated for this event (see Table \ref{tab:sample}).
For these parameters the late radio observations are constraining the maximal energy of the ejecta  to be $<10^{52}$ erg, provided that the mass is $\le 0.014\,\Msun$ (orange curve). The observations are much less constraining if the ejected mass is larger and if $\epb$ is smaller  (see Figure \ref{fig:const} below). 
For the case $\betain=0.3$ (blue), the narrow solid curves show how the light curve changes for different densities. In this case, for $n=2 \times 10^{-1} {\rm cm}^{-3}$, the radio upper limit constrains the maximal energy to $<10^{52}$~erg for ejecta masses $\le 0.12\,\Msun$.

The parallel vertical branches of the light curves describe the quenching due to the jet blast-wave \citep{Margalit&Piran2020}. The dashed branch describes the light curve if this quenching did not take place. We find that quenching is unimportant for these parameters. For large ejecta velocities, quenching is suppressed at earlier times, well before our observations took place. For lower ejecta velocities, the upper limit lies significantly above the predicted light curves and is unconstraining. 
This situation is typical for all events discussed here. 

\begin{figure*}
\begin{center}
\includegraphics[width=2.1\columnwidth]{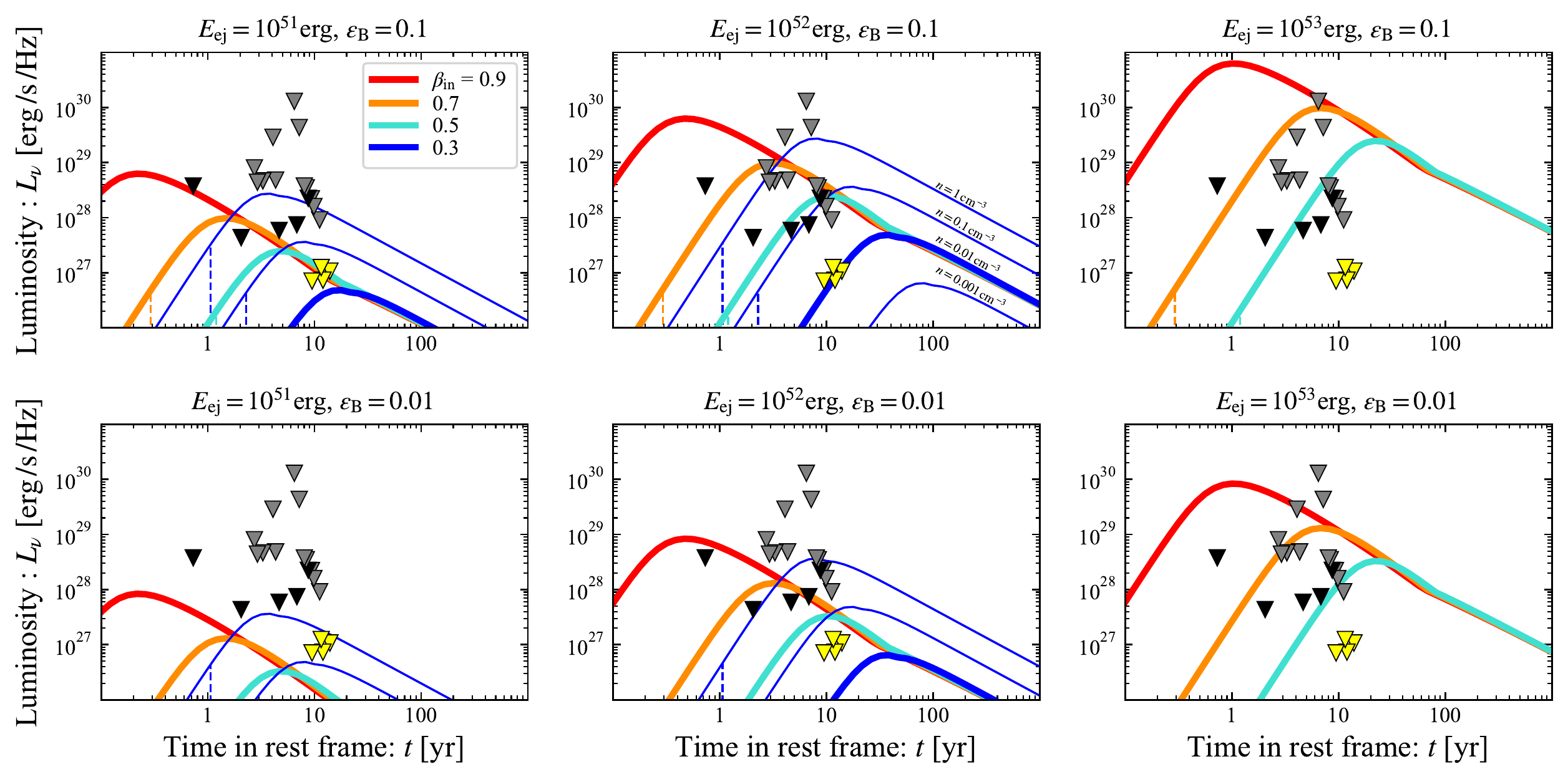}
\vspace{-0.4cm}
\caption{Constraints to the radio remnants of short GRBs. 
We report the observed upper limits for the sample of candidate local events (yellow symbols) as well as cosmological short GRBs at $z\approx$0.1 (black symbols) and above (grey symbols). 
Light curves expressed in term of the luminosity (6 GHz) for various initial velocities for $\Eej=10^{51}$ (left),  $10^{52}$ (middle), and $10^{53}$~erg (right)
 and $\epb=0.1$ (top) and $0.01$ (bottom). 
The tick solid curves assume a ISM density $n=10^{-2} {\rm cm}^{-3}$, other parameters are: $p=2.5$, $\bar{\varepsilon}_{\rm e}=0.1$.
The dashed vertical branch describes the quenching due to a jet (with $\Ej=10^{49} $ erg) blast-wave.}
\label{fig:lumi}
\end{center}
\end{figure*}

 Turning now to the whole sample we depict the observed upper limits on the luminosities (instead of observed fluxes) of all events in Figure \ref{fig:lumi}. 
This is done using the measured redshifts of the GRB host galaxies and the corresponding luminosity distances.  The luminosities of the candidate local group - GRBs 050906, 070810B, 080121, and 100216A - were estimated assuming a distance of 130 - 200 Mpc (see Table \ref{tab:vlasample}).
The observations are compared to a canonical light curve for an ejecta with $\Eej=10^{51}, 10^{52}$, and $10^{53}$~erg and for $\epb=0.1$ and $0.01$. For each combination of energy and magnetic field equipartition factor we consider different ejecta velocities that correspond to different masses (ranging from $\betain=0.3$ for $\Mej\simeq 0.012\,\Msun\,\Eejn$ to $\betain=0.9$ for $\Mej\simeq4.3\times10^{-4}\,\Msun\,\Eejn$).

\begin{figure*}
\begin{center}
\includegraphics[width=180mm, trim=40 0 5 5, clip]{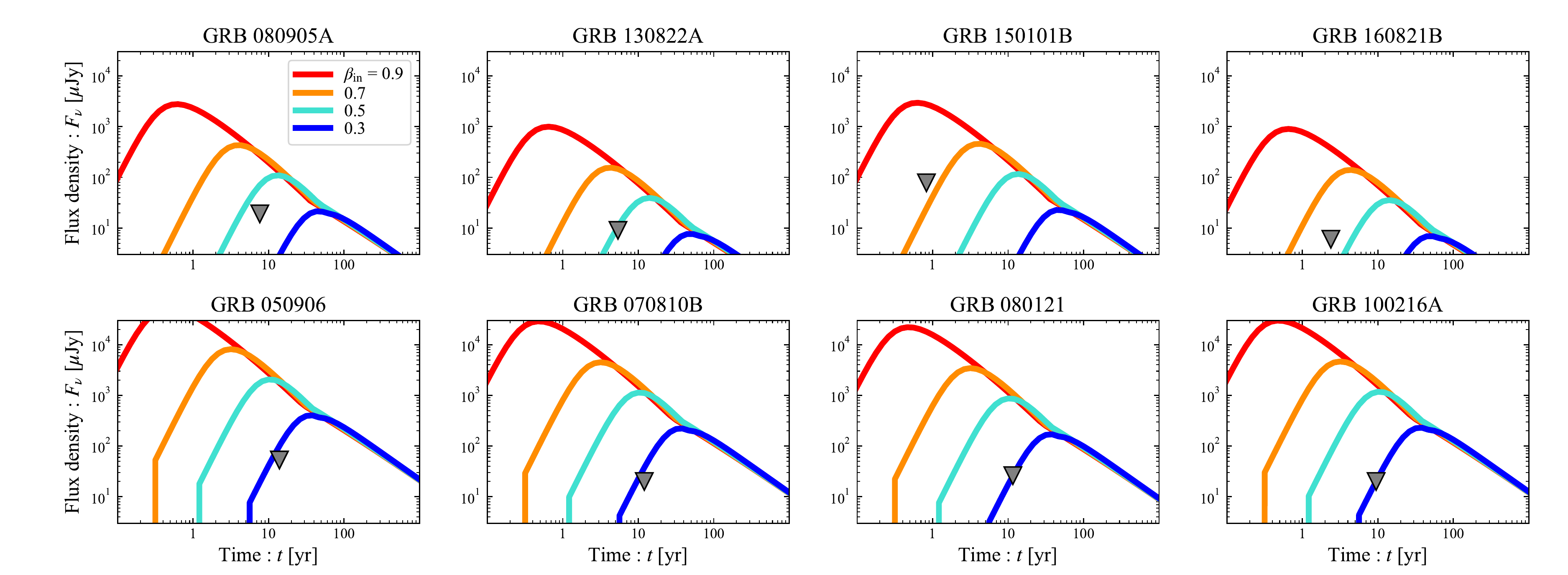}
\vspace{-0.4cm}
\caption{Radio light curves for $\Eej=10^{52}$ erg, $n=10^{-2}\,\rm cm^{-3}$, $\epb=0.1$, and different velocities, corresponding to ejecta masses of $\Mej\simeq0.004$\,$\Msun$ (red), $0.014$\,$\Msun$ (orange), $0.036$\,$\Msun$ (cyan) and $0.12$\,$\Msun$ (blue). The other parameters are $p=2.5$ and $\bar{\varepsilon}_{\rm e}=0.1$.
The top panels show cosmological events with $z\lesssim0.1$ and the bottom panels show candidate local ($<$200~Mpc) events.
The gray triangle shows the measured upper limit, where we report only the most constraining value if several observations are available.}
\label{fig:lc2}
\end{center}
\end{figure*}

In agreement with previous studies, we find that the most energetic scenario
(\Eej = $10^{53}$~erg) is disfavored in all cases considered, and would require a combination of low density ($<$0.01\,cm$^{-3}$) and low $\epb$ ($<$0.01) to be consistent with the observational constraints. 
For a circumburst density $n\approx10^{-2}\,{\rm cm}^{-3}$ and $\epb > 0.01$ the observations imply a significant limit on the energy deposited into the ejecta 
for practically all nearby ($z\approx0.1$) mergers (black symbols). 
Noticeably, the limits on the candidate local events (yellow symbols) are  more stringent than those found for the cosmological sample. With similar fluxes and significantly smaller distances,  naturally, the corresponding luminosities are much smaller leading to the stronger limits. Figure \ref{fig:lc2} focuses on the sample of nearby events:  the top panels show the limits for four sGRBs with $z\approx 0.1$ which place tighter constraints than the other cosmological events; the bottom panels show the limits for the four sGRBs that are possibly located within 200 Mpc.  
We fix the ejecta energy to $\Eej=10^{52}\,\rm erg$ as in Figure \ref{fig:lc} and the  ISM density to $n=10^{-2}\,\rm cm^{-3}$, which is a consistent value for all events (see Table \ref{tab:sample}). The other parameters are set to the same values as those in Figure \ref{fig:lc}. One can see that the current radio limits are significant (for this ejecta energy and ISM density) for these nearby events. With these chosen parameters, the observations  limit the ejecta velocity to less than $\beta_{\rm in}\lesssim0.3$ (or the ejected mass to be larger than $0.1\ \Msun$).

The upper limits on the energy are summarized in 
Figure~\ref{fig:const}. For each event, we show how the limits change as a function
of the circumburst density $n$. We consider a typical range $10^{-4}$\,cm$^{-3}$\,$\lesssim$\,$n$\,$\lesssim$1\,cm$^{-3}$, and show only the values consistent with the afterglow constraints (Table~\ref{tab:sample}). 
The region above the solid curve is ruled out by the observations. 
As the energy limits in this figure are a function of the ejecta mass we also highlight the relevant range of $0.01-0.05\,\Msun$. 

When the observation time is earlier than the deceleration time (namely, before the light curve peak), the flux upper limit imposes an upper limit on the velocity (see Eq. \eqref{eq:flux} for $t<\tdec$). 
This, in turn, implies that the maximal energy depends on the ejecta mass  increasing  linearly with it (the upper right part of each curve in Figure \ref{fig:const}).
When the observation time is later than the peak time, the light curves is determined just by the total energy and hence directly constrains it (the flat left side of each curve in Figure \ref{fig:const}). 
The transition between the two is when the observed time equals the peak. 
This is given by solving $t_{\rm obs}=\tdec$ and $F_{\rm obs}=F_{\nu}(\tdec)$, where $t_{\rm obs}$ and $F_{\rm obs}$ are the observation time and flux upper limit, respectively. We find the kinetic energy and ejecta mass at the transition point depends on the density and observables as $\Eej\propto n^{\frac{5p-19}{2(5p+13)}}t_{\rm obs}^{\frac{3(5p-7)}{5p+13}}F_{\rm obs}^{\frac{10}{5p+13}}$ and $\Mej\propto n^{\frac{7p-1}{2(5p+13)}}t_{\rm obs}^{\frac{3(5p+1)}{5p+13}}F_{\rm obs}^{\frac{6}{5p+13}}$, respectively. For most of our bursts, we find that $\Mej \lesssim 10^{-2}\,\Msun$ for $t_{\rm obs}\,\gtrsim\,\tdec$ when $\Eej$ is independent of $\Mej$. However, for some recent events, such as GRB~150101B and GRB~160821B, 
 the transition point is at $<10^{-3}\,\Msun$ and the allowed region is dictated by the lines of $\betain=\rm const$ in Figure \ref{fig:const}, which linearly increase with the ejecta mass. Future monitoring of these nearby events could lead to tighter constraints. This would be particularly interesting in the case of GRB~160821B, 
whose X-ray emission displays the typical plateau+steep decay profile, commonly interpreted as evidence for a magnetar \citep{Lu2017,Troja2019}. 
Optical/nIR observations suggest that a substantial mass of neutron-rich material was ejected from these systems \citep{Troja2018,Troja2019,Lamb2019}, 
which therefore represent promising candidates for future radio follow-up.

For $\Mej \lesssim 10^{-2}\,\Msun$ and other canonical parameters
we find that for almost all events a magnetar powering the ejecta with energy of $5\times 10^{52} $ ergs is ruled out, in agreement with past studies \citep{Horesh2016,Fong2016, Klose2019}. 
An exception is GRB 100206A provided that the surrounding ISM densities is at the minimal allowed value.   Not surprisingly  this is one of the furthest bursts ($z=0.4068$)  and hence the observational limit is less constraining. 
In some cases the limits are even stronger and constrain the ejecta energy to $\lesssim10^{52}$ erg or even lower values . In particular, this is the case for the group of candidate local events \citep{Dichiara2020}. 
The limits are somewhat relaxed if we consider the higher mass estimate,  $\Mej = 5 \times 10^{-2}\,\Msun$, but even with this large value an energy of $10^{53}$~erg is ruled out.

\begin{figure*}
\begin{center}
\includegraphics[width=180mm, trim=30 0 30 0, clip]{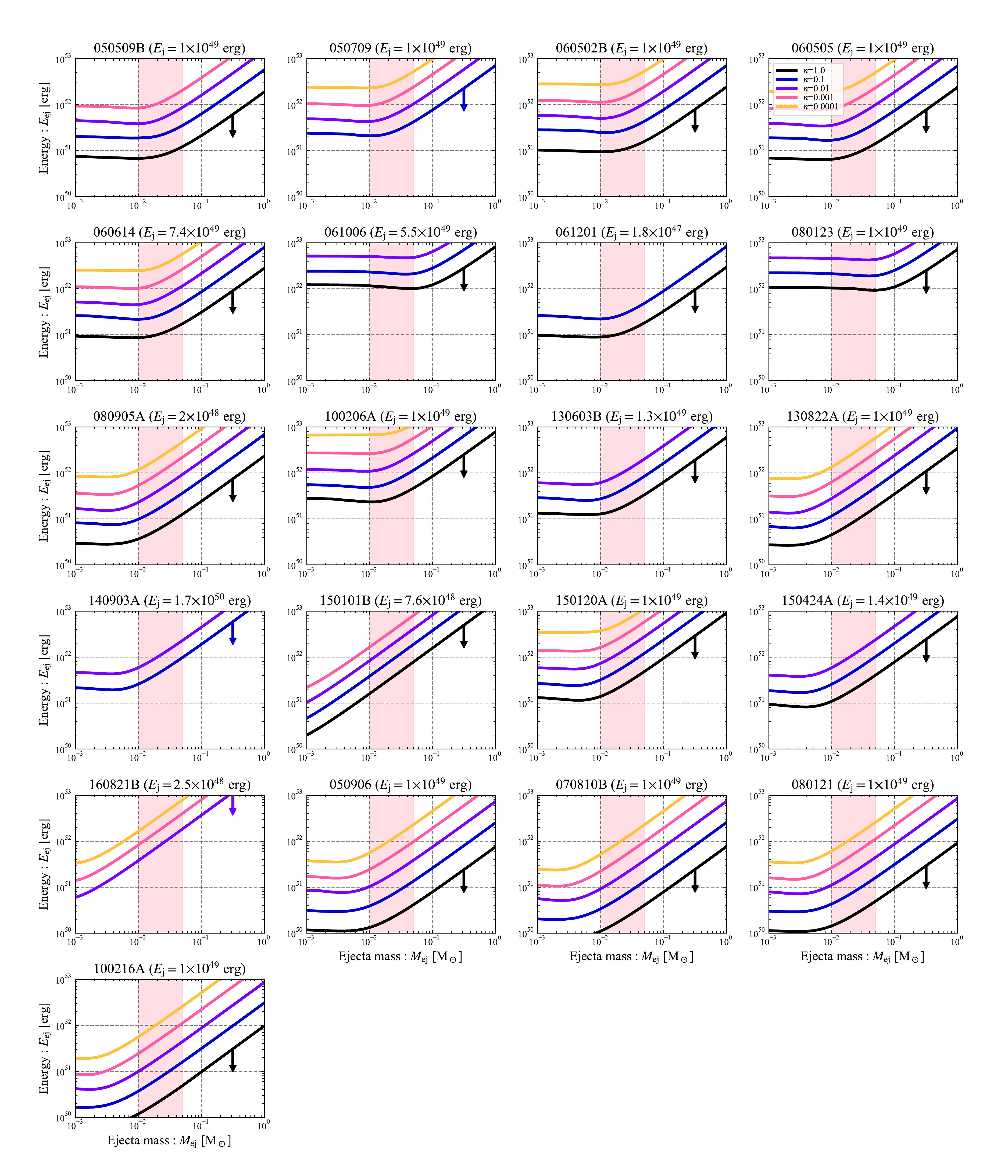}
\caption{Constraint on the ejecta mass and energy of each sGRB calculated for the parameters of $p=2.5$, $\bar{\varepsilon}_{\rm e}=0.1$, and $\varepsilon_{\rm B}=0.1$. We consider a range of the ISM density of $n=10^{-4}-1\,\rm cm^{-3}$. For each burst we show only densities that are consistent with the limits obtained with afterglow observations (see Table \ref{tab:sample}).
 Pink shaded regions show the ejecta mass range $0.01<\Mej/\Msun<0.05$.
} 

\label{fig:const}
\end{center}
\end{figure*}

\begin{figure}
\begin{center}
\includegraphics[width=85mm, angle=0]{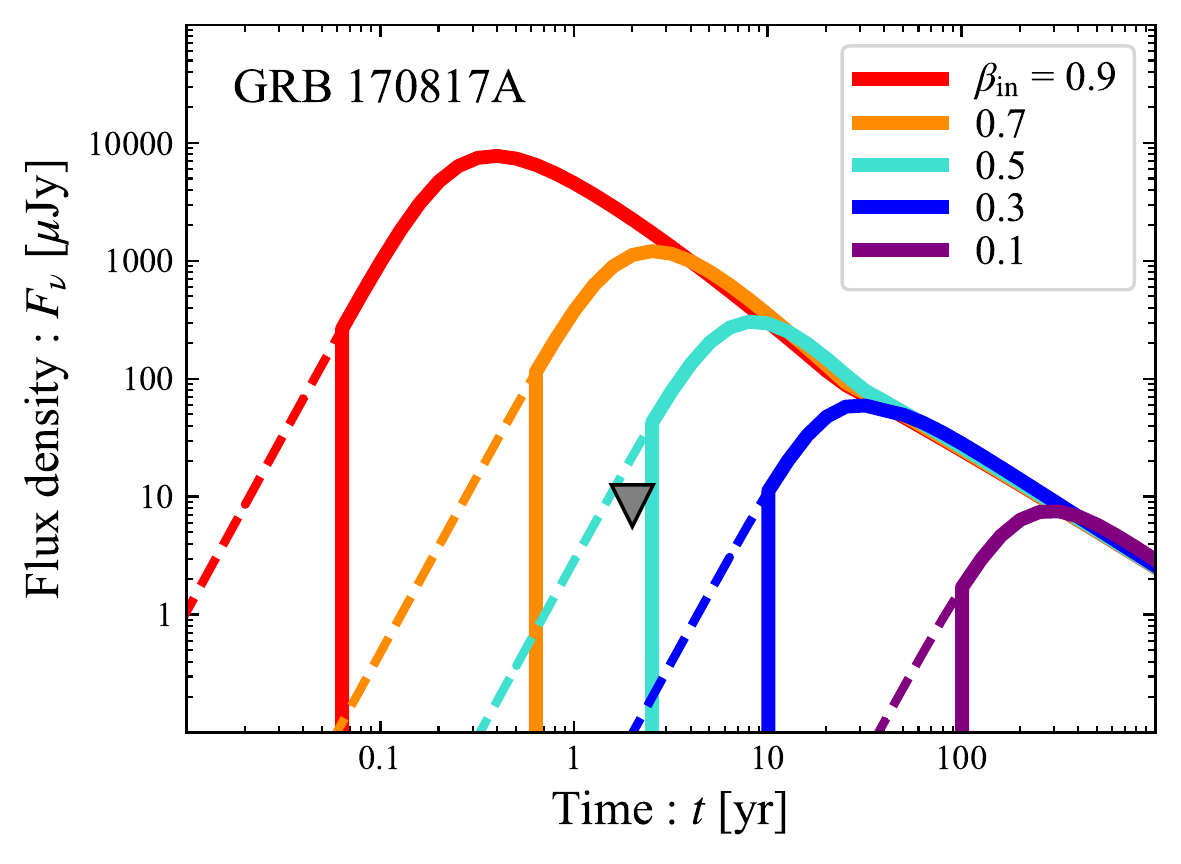}
\caption{The same as Figure \ref{fig:lc} but for GRB 170817A.
The adopted parameters are $\Eej=10^{51}\,\rm erg$, $n=2\times10^{-3}\,\rm cm^{-3}$, $\varepsilon_{\rm B}=0.1$, $p=2.5$, $\bar{\varepsilon_{\rm e}}=0.1$, and $E_{\rm j}=10^{49}\,\rm erg$.}
\label{fig:lc170817}
\end{center}
\end{figure}

\begin{figure}
\begin{center}
\includegraphics[width=85mm, angle=0]{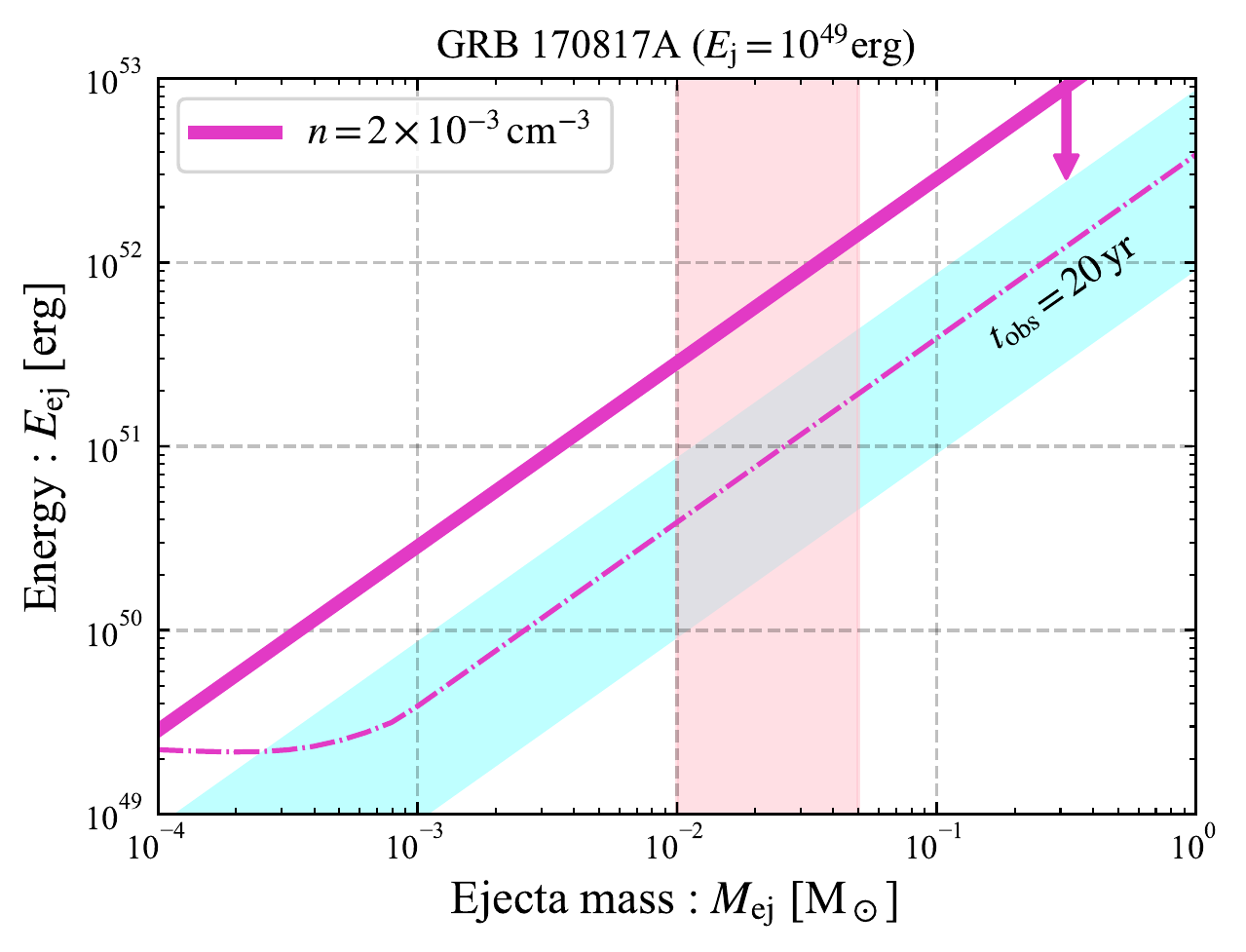}
\caption{The same as Figure \ref{fig:const} but for GRB 170817A. The pink and cyan shaded regions show the range of ejecta mass $\Mej=0.01-0.05\,\Msun$ and velocity $\betain=0.1-0.3$, respectively. The dash-dotted curve demonstrates how we can reduce the allowed region by late-time observations at 20 yr after the merger, assuming a ten-fold improvement in sensitivity \citep[e.g.][]{butler18}. }
\label{fig:const170817}
\end{center}
\end{figure}

For completeness we add in Figures \ref{fig:lc170817} and \ref{fig:const170817}  the light curve and the ejecta energy constraint for GRB 170817A. 
Unlike other sGRBs, the quenching by the jet can be relevant for this event \citep[see $\betain=0.3$ curve in Figure \ref{fig:lc170817} and][]{Margalit&Piran2020}, and does not allow us to constrain $\Eej$ by the flux upper limit. 
Instead, the condition $t_{\rm obs}<t_{\rm col}$ constrains the maximal velocity. The allowed region on the $\Eej$-$\Mej$ plane (shown in Figure \ref{fig:const170817}) is below the line corresponding to  $\betain=\rm const$ (magenta curves).
While this event occurred at much closer distance than any other sGRB ($\approx$40~Mpc), the current dataset yield constraints similar to other events due to the short time elapsed from the NS merger ($\lesssim$2.5 yrs). 
Only continued monitoring of this source could provide tighter limits on the ejecta energy: Figure \ref{fig:const170817} shows the expected limit at 20 yrs post-merger provided a 10 times deeper upper limit flux. Clearly this would provide a much more stringent constraint. 

\section{Conclusions} \label{sec:conclusion}

We present a systematic radio survey of sGRB locations searching for any possible late-time rebrightening.  As suggested by \cite{np11}, merger ejecta sweep up the circumburst medium and may produce a delayed radio flare as seen in supernovae \citep{Chevalier1982,Chevalier1998}.
If a rapidly spinning magnetar is formed after the NS merger, it will inject energy into the ejecta, which could result in a brighter radio emission.
We therefore use the derived radio limits to constrain the energy of merger ejecta, and test the presence of a magnetar central engine. 

Our sample includes 17 GRBs at cosmological distance with $z$\,$\lesssim$0.5, 
4 GRBs possibly associated to nearby galaxies (100~Mpc$\lesssim$\,$d$\,$\lesssim$200~Mpc), and the gravitational wave event GRB~170817A ($d$\,$\approx$40 Mpc). 
For the majority of these bursts ($\sim$60\% of our sample) the formation of a long-lived NS was suggested in order to explain the peculiar temporal features of their X-ray afterglows (Table~\ref{tab:sample}). In some of these events, evidence for a substantial ($\gtrsim$0.01\,$M_{\odot}$) mass ejection is supported by the identification of possible kilonova emission
\citep{Tanvir2013,Yang2015, Jin16, Troja2018,Troja2019,Lamb2019}. 
In other cases, namely GRB~080905A, GRB~050509B and the candidate local GRBs, 
deep optical limits suggest  $\lesssim$\,0.01\,$M_{\odot}$ \citep{Gompertz2018,Dichiara2020}.

We find no radio counterpart to any of these events, down to a typical luminosity 
of $\nu L_{\nu}$\,$\approx$\,(5-50)$\times$10$^{37}$\,erg\,s$^{-1}$ for cosmological ($z\gtrsim$0.1) events, and of $\nu L_{\nu}$\,$\approx$\,5$\times$10$^{36}$\,erg\,s$^{-1}$ for local ($\lesssim$200~Mpc) events. 
Our work improves upon the constraints obtained by \citet{Fong2016} and \citet{Liu20} in several ways. 
We focus on the closest events: the sample considered in these past works has a median distance of $\approx$2.5~Gpc, while our cosmological sample lies at $\approx$1.1~Gpc and, in addition, our study includes several events possibly located at $\lesssim$200~Mpc. Thanks to the lower distance scale and higher sensitivity of our observations, we can probe luminosities that are an order of magnitude lower. 
Our observations span a wide range of timescales,  from $\approx$0.8~yr for GRB~150101B to $\approx$14~yr for GRB~050906, with a median value of $\approx$9~yr after the merger. This is a factor of 2 longer than the timescales considered in \citet{Liu20} and \citet{Fong2016}, and allows us to probe lower velocities and larger ejecta masses.

By comparing the theoretical light curves with the flux upper limits of these 22 GRBs, we constrained the allowed range of ejecta mass and energy (see Figures \ref{fig:const} and \ref{fig:const170817}).
For an ejecta mass $\Mej$\,$\lesssim$\,0.01\,$\Msun$, expected for typical mergers \citep[e.g.][]{Bauswein13,Hotokezaka13},  we can disfavor energetic ejecta $\Eej$\,$\gtrsim$\,5$\times$10$^{52}$\,erg under reasonable assumptions on the microphysical equipartition parameters (\epb$\gtrsim$0.01-0.1, \bepe$\approx$0.1). For higher ejecta mass, up to $\Mej\lesssim 0.05\,\Msun$, only energies $\Eej\gtrsim10^{53}$\,erg can be ruled out.  
Tighter constraints can be derived  by taking into account the range of densities
allowed by afterglow observations (Figure~\ref{fig:const}). For $\Mej\lesssim 0.05\,\Msun$ we can rule out $\Eej$\,$\gtrsim$\,(2-5)$\times$10$^{52}$erg for 11 events (65\% of the cosmological sample) and $\Eej\gtrsim5\times10^{51}\rm erg$ for one event.

Similar constraints are derived for the group of candidate local sGRBs \citep{Dichiara2020}. Despite their closer distances, these events have 
only mild constraints on their circumburst density and, by considering values as low as $n\gtrsim 10^{-4}$cm $^{-3}$, we can confidently rule out $\Eej$\,$\gtrsim$\,2$\times10^{52}$\,erg. Lower ejecta masses ($\lesssim$0.01\,\Msun) would explain the non-detection of a kilonova and imply more stringent limits  $\Eej$\,$\gtrsim$\,6$\times10^{51}$\,erg.

A limit $\Eej$\,$\gtrsim$\,2$\times10^{52}$\,erg is derived also for GRB~170817A. In this case, the available observations probe an early phase of the evolution, and we find that quenching of the radio emission by the GRB jet \citep{Margalit&Piran2020} might be significant at this stage. Long-term radio monitoring of this source could substantially improve upon these limits (Figure~\ref{fig:const170817}). 
We caution however that, if powered by radioactive decay the total ejecta mass might be slightly higher than 0.06\Msun within the uncertainties of the modelling (i.e.,the three-component ejecta model in \citet{Villar17}) and this slightly increases the upper bound on $\Eej$. On the other hand, a long-lived Neutron Star or long-lasting BH activity can provide additional energy to power the kilonova and in this case the required mass of the merger ejecta could be somewhat smaller than that required by the radioactive power model \citep{Yu2018,Matsumoto18}.

The range of energies probed by our observations are comparable to the rotational energy of a rapidly spinning NS, that can be as high as $E_{\rm rot}$\,$\approx$\,$10^{53}$\,erg for a heavy magnetar with
a spin period  $P$\,$\approx$\,0.7~{\rm ms}.
Such rotational energy can be  effectively transferred to the ejecta through magnetic dipole radiation.  On the other hand, in a newly formed NS gravitational wave losses can tap away rotational energy much more effectively than electromagnetic losses \citep{Dallosso15},  reducing the energy dumped into ejecta and thus allowing for a much larger rotational energy. 

The radio limits are also consistent with the range of energies required by X-ray observations. X-ray plateaus, with typical observed luminosities of $L_X$\,$\approx$10$^{47}$-10$^{48}$\,erg\,s$^{-1}$ and durations $T$\,$\approx$\,100-1,000\,s, imply a spin-down 
energy $\lesssim$10$^{52}$\,erg for a radiative efficiency $\eta_X$\,$\gtrsim$0.01. 
A comparable upper bound is derived by sGRBs with temporally extended emission, except for the extreme case of GRB~060614 for which the measured energy during the EE phase is already 3$\times$10$^{51}$\,erg \citep{Gompertz2013}, 
and the total energy release is estimated as $E_\textrm{iso}$\,$\approx$2$\times$10$^{52}$\,erg \citep{Lu2015}. 
A narrow beaming factor could easily reconcile these values with the limits imposed by radio observations. 

The lack of a long-lived remnant in the majority of sGRBs may indicate that a sizable fraction of binary NS mergers lead to a BH remnant 
\citep{Piro2017}, thus favoring soft NS equations of state, or provide further evidence for a different channel of progenitor systems, 
such as NS-BH mergers \citep{Troja2008, Gompertz2020, Thakur2020, Fernandez20}.
Our results indicate that if energetic ejecta with $\Eej\gtrsim 10^{52}\rm\,erg$ are present, as suggested by \citet{Liu20}, the radio flux is expected to rise above the current sensitivity limit in the sample of close ($z\lesssim 0.1$) events (Figures \ref{fig:lc2} and \ref{fig:lc170817}). Future observations, on a time scale of a few years, will be then crucial to disclose the emergence of this component or further constrain the energy of the ejecta.

\section*{Acknowledgements}

We thank the anonymous referee for their useful comments.
We acknowledge helpful discussions with Paz Beniamini  and Ben Margalit. This work made use of data supplied by the UK \textit{Swift} Science Data Centre at the University of Leicester. GB acknowledges financial support under the INTEGRAL ASI-INAF agreement 2019-35-HH.0, TP acknowledges support by an ERC advanced grant TReX. TM is supported by JSPS Postdoctral Fellowship, Kakenhi No. 19J00214. BO is supported in part by the National Aeronautics and Space Administration through grants NNX16AB66G, NNX17AB18G, and 80NSSC20K0389. AC acknowledges support from NSF grant 1907975. The National Radio Astronomy Observatory is a facility of the National Science Foundation operated under cooperative agreement by Associated Universities, Inc.  

\section*{Data availability}
The data underlying this article will be shared on reasonable request to the corresponding author.




\bibliographystyle{mnras}









\bsp	
\label{lastpage}
\end{document}